\documentclass[aps,prl, reprint, 
showpacs,showkeys,superscriptaddress,preprintnumbers,nofootinbib]{revtex4-1} 

\usepackage{amsmath,amssymb,bm,mathrsfs}
\usepackage{hyperref}
\usepackage{url}
\usepackage{graphicx}

\begin{document}


\title{Axion Quality Problem Alleviated by Non-Minimal Coupling to Gravity} 

\author{Koichi Hamaguchi}
\email{hama@hep-th.phys.s.u-tokyo.ac.jp}
\affiliation{Department of Physics, University of Tokyo, 
Tokyo 113--0033,
Japan}
\affiliation{
Kavli Institute for the Physics and Mathematics of the Universe (Kavli
IPMU), University of Tokyo, Kashiwa 277--8583, Japan
}

\author{Yoshiki Kanazawa}
\email{kanazawa@hep-th.phys.s.u-tokyo.ac.jp}
\affiliation{Department of Physics, University of Tokyo, 
Tokyo 113--0033,
Japan}

\author{Natsumi Nagata}
\email{natsumi@hep-th.phys.s.u-tokyo.ac.jp}
\affiliation{Department of Physics, University of Tokyo, 
Tokyo 113--0033,
Japan}

\begin{abstract}

  Global symmetries are expected to be violated by gravity, which may cause a serious problem to models based on these symmetries. A famous example is the Peccei-Quinn solution to the strong CP problem in QCD axion models; it requires a global U(1) symmetry to be respected with high quality, and even the Planck-scale suppressed operators that violate the U(1) symmetry would spoil this solution. Indeed, it is known that gravitational instantons do break the U(1) Peccei-Quinn symmetry and induce the U(1)-violating operators, bringing back the strong CP problem to the axion models. This conclusion is, however, dependent highly on the structure of the theory around the Planck scale and, therefore, may be evaded if we go beyond the minimal setup. In this paper, we study the effect of a non-minimal coupling of the Peccei-Quinn field to gravity, $\xi$, on the gravitational instanton effect. This setup is frequently considered in cosmology as it can realize a successful inflation if $\xi \lesssim 10^5$. We find that the U(1)-breaking effect of the gravitational instantons can sufficiently be suppressed for $\xi \gtrsim 2 \times 10^3$, which suggests that the U(1) symmetry may be maintained with high quality even in the presence of gravity. Our result thus points to a new way to avoid the quality problem of global symmetries. 

\end{abstract}

\maketitle

\section{Introduction}

Global symmetries play crucial roles in many models beyond the Standard Model. A prominent example is the Peccei-Quinn model~\cite{Peccei:1977hh,Peccei:1977ur}, where an anomalous global U(1) symmetry is introduced to solve the strong CP problem. After this U(1) symmetry is spontaneously broken, a Nambu-Goldstone boson, called axion~\cite{Weinberg:1977ma,Wilczek:1977pj}, appears in the low-energy effective theory. Once QCD is confined, a periodic scalar potential for the axion field is generated non-perturbatively and the axion field relaxes to a CP-preserving vacuum, solving the strong CP problem. See Refs.~\cite{DiLuzio:2020wdo, Sikivie:2020zpn, Agrawal:2021dbo} for recent reviews on axion models.

For the Peccei-Quinn mechanism to solve the strong CP problem, any effect that explicitly violates the U(1) symmetry should be smaller than the non-perturbative QCD effect; otherwise, the minimum of the axion potential in general deviates from the CP-conserving points. On the other hand, any global symmetries are expected to be violated by gravity~\cite{Banks:2010zn, Witten:2017hdv, Harlow:2018jwu, Harlow:2018tng}, and hence there exists some amount of explicit violation of the Peccei-Quinn symmetry. To make matters worse, it is discussed based on the effective theoretical approach in Refs.~\cite{Georgi:1981pu, Dine:1986bg, Kamionkowski:1992mf, Holman:1992us, Barr:1992qq, Ghigna:1992iv} that this U(1)-violating effect is generically too large to solve the strong CP problem due to the high sensitivity of the Peccei-Quinn mechanism to the U(1)-breaking effect---this is dubbed as the axion quality problem. This observation stimulated a variety of works to construct axion models that evade the quality problem, such as those utilizing additional gauge symmetries~\cite{Chun:1992bn, Cheng:2001ys, Hill:2002kq, Babu:2002ic, Dias:2002gg, Harigaya:2013vja, Fukuda:2017ylt, DiLuzio:2017tjx, Duerr:2017amf, Fukuda:2018oco, Bonnefoy:2018ibr, Ibe:2018hir, Choi:2020vgb, Ardu:2020qmo, Yin:2020dfn, DiLuzio:2020qio, Kawamura:2020jzb, Darme:2021cxx, Chen:2021haa}, heavy axion models~\cite{Dimopoulos:1979pp, Tye:1981zy, Rubakov:1997vp, Berezhiani:2000gh, Gianfagna:2004je, Hook:2014cda, Fukuda:2015ana, Albaid:2015axa, Chiang:2016eav, Gherghetta:2016fhp, Dimopoulos:2016lvn, Agrawal:2017ksf, Gherghetta:2020keg, Gherghetta:2020ofz}, composite axion models~\cite{Kim:1984pt, Choi:1985cb, Randall:1992ut, Redi:2016esr, Lillard:2017cwx, Lillard:2018fdt, Lee:2018yak, Gavela:2018paw, Ishida:2021avk}, and so on~\cite{Carpenter:2009zs, Yamada:2015waa,Higaki:2016yqk, Cox:2019rro, Bonnefoy:2020llz,Yamada:2021uze, DiLuzio:2021pxd,Nakai:2021nyf, Bhattiprolu:2021rrj}. 

It is rather difficult to explicitly evaluate the extent of the gravitational U(1)-breaking, since this is essentially a non-perturbative quantum effect. There is, however, a class of processes which we can analyze quantitatively---gravitational instantons or, in other words, Euclidean wormholes. Euclidean wormholes are configurations of gravitational fields in the Euclidean spacetime that connect two asymptotically flat spacetime regions through a throat. In Ref.~\cite{Giddings:1987cg}, S.~B.~Giddings and A.~Strominger found Euclidean wormhole solutions in the dual version of the axion theory described by a two-form gauge field. This setup corresponds to the model that contains only the axion degree of freedom, not a radial component as in the Peccei-Quinn model. Later, it was demonstrated in Ref.~\cite{Lee:1988ge} that such wormhole solutions also exist in theories with a complex scalar field that possess a U(1) symmetry, as in standard axion models. These wormholes indeed induce the U(1)-violating effects, and these effects can be expressed in terms of effective local operators that break the U(1) symmetry~\cite{Rey:1989mg, Abbott:1989jw, Coleman:1989zu}. For recent studies on axionic wormholes, see Refs.~\cite{Hebecker:2016dsw, Alonso:2017avz, Hebecker:2018ofv, Alvey:2020nyh}. 

The coefficients of the U(1)-violating effective operators induced by wormholes turn out to be highly model-dependent~\cite{Kallosh:1995hi}; since they are generated in non-perturbative processes, the coefficients are proportional to the factor $e^{-S}$, where $S$ denotes the wormhole action, and thus a sizable value of $S$ can considerably suppress these operators. In fact, it is found that in the case of the Giddings-Strominger wormholes~\cite{Giddings:1987cg}, the wormhole action is large enough to sufficiently suppress the U(1)-violating operators if the axion decay constant is $f_a \lesssim 10^{16}$~GeV~\cite{Alonso:2017avz}. The situation, however, drastically changes if the axion is a part of a complex scalar field and couples minimally to gravity; in this case, the wormhole action reduces due to the presence of the radial component of the scalar field and, as a consequence, the wormhole contribution to the U(1) violation is sizable and spoils the Peccei-Quinn solution to the strong CP problem for any allowed values of $f_a$~\cite{Kallosh:1995hi, Alvey:2020nyh}. These two examples imply that the extent of the gravitational U(1)-breaking effect strongly depends on the ultraviolet completion of the axion degree of freedom, which motivates us to explore models beyond the minimal setup. 

In this work, we consider the case where the Peccei-Quinn complex scalar field has a non-minimal coupling to gravity, $\xi$. 
A similar setup has, in fact, been suggested in cosmology~\cite{Futamase:1987ua, Salopek:1988qh, Fakir:1990eg, Makino:1991sg, Fakir:1992cg, Barvinsky:1994hx, Kaiser:1994vs, Kamenshchik:1995ib, Mukaigawa:1997nh, Libanov:1998wg, Komatsu:1999mt}, as it can realize a successful inflation if $\xi \lesssim 10^5$. See Refs. \cite{Linde:2011nh, Kaiser:2013sna, Fairbairn:2014zta, Ballesteros:2016euj, Ballesteros:2016xej, Boucenna:2017fna, McDonough:2020gmn} for recent studies on inflation models with a non-minimal coupling to gravity.
We study the wormhole solution in the presence of this non-minimal coupling~\cite{Coule:1989xu, Coule:1992pz}, and the effect of it on the wormhole action. Our main finding is that the resultant U(1)-breaking effect can sufficiently be suppressed for $\xi \gtrsim 2 \times 10^3$.
This model can thus offer an inflation that is compatible with the observed CMB data~\cite{Planck:2018jri}, while saving the axion from dangerous U(1)-violating effects.

\section{Model}

We consider a theory of a complex scalar field $\Phi (x)$ in the Euclidean spacetime. We assume that the metric has the spherically symmetric form: $ds^2  = dr^2 + a(r)^2 d^2 \Omega_3$, where $a(r)$ is the scale factor and $d^2 \Omega_3$ is the line element on the three-dimensional sphere. The complex scalar field $\Phi (x)$ is expressed in terms of two real scalar fields, $f(x)$ and $\theta (x)$, as $\Phi (x) = f(x)e^{i \theta (x)}/\sqrt{2}$. These fields are assumed to depend only on $r$ in accordance with the spherically symmetric metric. 

The field $\Phi$ is supposed to have a non-minimal coupling to the Ricci scalar $R$, as described by the following action: $S_{\mathrm{tot}} = S + S_{\mathrm{GHY}}$, with 
\begin{align}
  S = \int d^4 x \, \sqrt{g} \biggl[- \frac{M^2}{2} R - \xi |\Phi|^2 R + |\partial_\mu \Phi|^2 + V(\Phi) \biggr] &~, \label{eq:act} \\ 
  V(\Phi) = \lambda \left( |\Phi|^2 - \frac{f_a^2}{2}\right)^2 &~,
  \label{eq:vphi}
\end{align}
where $g \equiv \mathrm{det} (g)$, $\xi$ is the non-minimal coupling, and $f_a$ is the axion decay constant. In the asymptotically flat spacetime regions, the field $\Phi$ develops a vacuum expectation value, $\langle |\Phi| \rangle = f_a/\sqrt{2}$. The mass parameter in Eq.~\eqref{eq:act}, $M$, is determined from the condition $M_P^2 = M^2 + \xi f_a^2$, with $M_P \equiv  (8\pi G)^{-1/2}$ the reduced Planck mass. We require $M^2 \geq 0$ so that the kinetic term of the gravitational field has the correct sign for arbitrary values of the field $\Phi$; this means $\xi \leq M_P^2/f_a^2$.\footnote{The case with $M = 0$, \textit{i.e.}, $\xi f_a^2 = M_P^2$, is sometimes called the induced-gravity model and has been studied in, \textit{e.g.}, Refs.~\cite{Zee:1978wi, Smolin:1979uz, Adler:1982ri, Spokoiny:1984bd, Accetta:1985du, Lucchin:1985ip, Salopek:1988qh, Fakir:1990iu, Kaiser:1994wj, Kaiser:1994vs, Cervantes-Cota:1995ehs}.} For later use, we define
\begin{equation}
  \Omega^2 (\Phi) \equiv 1 + \frac{2 \xi}{M_P^2} \left( |\Phi|^2 - \frac{f_a^2}{2}\right) ~.
\end{equation} 
$S_{\mathrm{GHY}}$ is the {Gibbons-Hawking-York (GHY)} term~\cite{York:1972sj, Gibbons:1976ue}: 
\begin{align}
  S_{\mathrm{GHY}} = - M_P^2 \int_{\partial V} d^3 x \,  \sqrt{\widetilde{g}} \, \Omega^2 (\Phi) (K - K_0) ~,
  \label{eq:sghy}
\end{align}
where $K$ is the extrinsic curvature for the boundary $\partial V$, $K_0$ is that for the boundary embedded into the flat spacetime, and $\widetilde{g}$ is the determinant of the induced metric on $\partial V$. The GHY term is added to make the variational problem well-defined. Since this is a surface term, it does not affect the equations of motion, though it does modify the value of the action. $K_0$ is subtracted from $K$ to make the action finite.

We compute the transition amplitudes for gravitational instanton processes using the Euclidean path-integral formalism in the semi-classical approximation. As discussed in Ref.~\cite{Coleman:1989zu}, there is a subtlety in this calculation; reflecting the invariance of the action under the global U(1) transformations, $\Phi (x)  \to e^{i \alpha} \Phi (x)$, the path integral is subject to the constraint of the form $\partial_\mu J^\mu = 0$ with $J^\mu = \sqrt{g}\, g^{\mu\nu} f^2 \partial_\nu \theta$, where $g^{\mu\nu}$ is the inverse of the metric $g_{\mu\nu}$. We must take account of this constraint when we search for stationary points. In practice, we can find correct stationary solutions by minimizing the following action with respect to $J_\mu$, $f$, and $g_{\mu \nu}$: 
\begin{align}
  S = \int d^4 x \, \sqrt{g} &\biggl[- \frac{M_P^2}{2} \Omega^2 (f) R + \frac{1}{2} (\partial_\mu f)^2  + V(f) \nonumber \\ 
  &+ \frac{1}{2g f^2} g_{\mu\nu} J^\mu J^\nu + \frac{1}{\sqrt{g}} \theta \partial_\mu J^\mu 
  \biggr] ~, 
\end{align}
where $\theta$ is introduced as a Lagrange multiplier, and the variation with respect to $J^\mu$ leads to the expression of $J^\mu$ mentioned above. From the condition $\partial_\mu J^\mu = 0$, we then find that the spatial integral of $J^0$ is constant, \textit{i.e.}, 
\begin{equation}
  2\pi^2 a^3(r) f^2 (r) \theta^\prime (r) = n ~,
  \label{eq:charge}
\end{equation}
where the prime represents the derivative with respect to $r$. Moreover, it turns out that $n$ is an integer; this follows from the periodicity of $\theta$: $\theta = \theta + 2\pi k$ for $k \in \mathbb{Z}$.\footnote{See, for instance, Ref.~\cite{Alonso:2017avz}.  } 

On the other hand, the variation with respect to $f$ gives 
\begin{align}
  f^{\prime \prime} + 3 \frac{a^\prime}{a} f^\prime = \frac{dV}{df} - \frac{n^2}{4\pi^4 f^3 a^6} + 6 \biggl[\frac{a^{\prime \prime}}{a} + \frac{a^{\prime 2}}{a^2} - \frac{1}{a^2}\biggr] \xi f ~,
  \label{eq:eomf}
\end{align}
while for the variations of the metric, we obtain
\begin{align}
  & \Omega^2 (f) \left( a^{\prime 2} -1 \right) + \frac{2 \xi }{M_P^2} a a^\prime f f^\prime \nonumber \\ 
  &= - \frac{a^2}{3 M_P^2} \biggl[- \frac{1}{2} f^{\prime 2} + V(f) + \frac{n^2}{8\pi^4 f^2 a^6} \biggr] ~, \label{eq:eommett} \\[3pt] 
  &\Omega^2 (f) \left(2 a a^{\prime \prime} + a^{\prime 2} -1 \right) + \frac{2 \xi a^2}{M_P^2} \biggl[f f^{\prime \prime} + f^{\prime 2} + 2 \frac{a^\prime}{a} f f^\prime \biggr] \nonumber \\
  &= - \frac{a^2}{M_P^2} \biggl[\frac{1}{2} f^{\prime 2} - \frac{n^2}{8 \pi^4 f^2 a^6} + V(f)\biggr] ~,
  \label{eq:eommetsp}
\end{align}
where we have used Eq.~\eqref{eq:charge} to eliminate $\theta^\prime$. It turns out that Eq.~\eqref{eq:eommetsp} follows from the other two equations, and therefore only Eq.~\eqref{eq:eomf} and Eq.~\eqref{eq:eommett} are the independent equations to be solved. Since Eq.~\eqref{eq:eomf} and Eq.~\eqref{eq:eommett} are second- and first-order differential equations, respectively, we need three boundary conditions in total, for which we use\footnote{As pointed out in Ref.~\cite{Alonso:2017avz}, in practice, it is more convenient to use a second-order differential equation for $a^{\prime \prime}$ obtained from Eq.~\eqref{eq:eommett} and Eq.~\eqref{eq:eommetsp}. In this case, an additional boundary condition for $a (0)$ is determined as a function of $f(0)$ from Eq.~\eqref{eq:eommett}. } 
\begin{align}
  f^\prime (0) = 0 ~, \quad f(\infty) = f_a ~, \quad a^\prime (0) = 0 ~.
  \label{eq:bcond}
\end{align}
Namely, we search for a ``half-wormhole'' configuration with a ``throat'' at $\rho \simeq 0$. To that end, we numerically solve the differential equations with the boundary conditions~\eqref{eq:bcond}.

\section{Result}

\begin{figure}
  {\includegraphics[clip, width = 0.45 \textwidth]{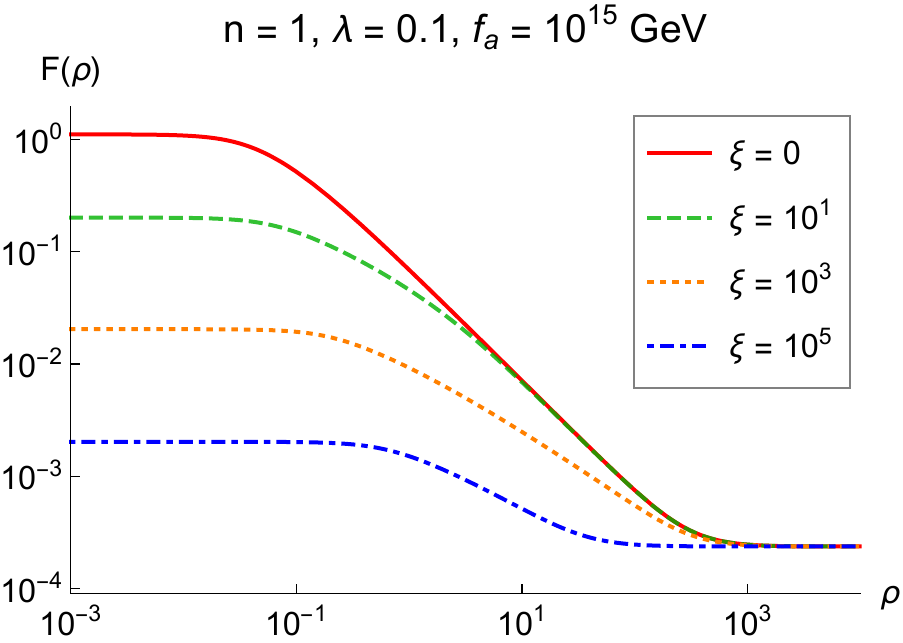}} 
  \caption{
  $F(\rho)$ for several values of $\xi$. 
  }
  \label{fig:F-rho}
  \end{figure}
\begin{figure}
  {\includegraphics[clip, width = 0.45 \textwidth]{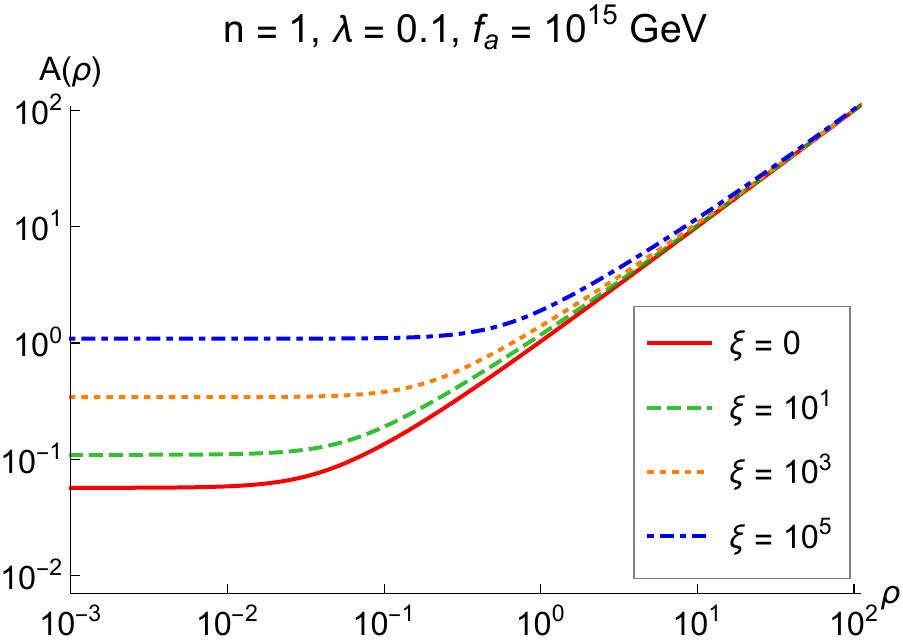}} 
  \caption{
    $A(\rho)$ for several values of $\xi$. 
  }
  \label{fig:A-rho}
  \end{figure}

Following Ref.~\cite{Abbott:1989jw}, we define the dimension-less quantities  
\begin{align}
  \rho \equiv \sqrt{3\lambda} \,   M_P r ~, \quad 
  A \equiv \sqrt{3\lambda} \, M_P a ~,\quad 
  F \equiv \frac{f}{\sqrt{3} M_P} ~.
  \label{eq:var}
\end{align}
In Fig.~\ref{fig:F-rho} and Fig.~\ref{fig:A-rho}, we show $F(\rho)$ and $A(\rho)$ for several values of $\xi$, respectively, where we set $n = 1$, $\lambda = 0.1$, and $f_a = 10^{15}$~GeV. As we see, the effect of the non-minimal coupling on the field configurations is considerable; the radial field $F(\rho)$ (scale factor $A(\rho)$) at $\rho = 0$ decreases (increases) as $\xi$ gets larger. In the asymptotic regions $\rho \to \infty$, $A (\rho) \to \rho$, which corresponds to the flat metric. 

In terms of the variables in Eq.~\eqref{eq:var}, the action is expressed as 
\begin{align}
  &S = \frac{2 \pi^2}{\lambda} \int_0^{\infty} d\rho \, 
  \left[ \Omega^2 (F) A^2 A^{\prime \prime} - 6 \xi A^2 A^\prime F F^\prime + A^3 F^{\prime 2} \right] ~, \\ 
  &S_{\mathrm{GHY}} = - \frac{2 \pi^2}{\lambda}  A^2 (0) \Omega^2 (F(0)) ~,
  \label{eq:sghyv}
\end{align}
where the prime denotes the derivative with respect to $\rho$.
As we see, the contribution of the GHY term comes only from the boundary at $\rho = 0$ and always reduces the total action. It is, however, not clear if we should really need to include this term~\cite{Kallosh:1995hi}. For instance, if we consider a ``full-wormhole'' configuration connecting two asymptotically-free regions, there is no boundary at the throat and thus the GHY contribution is absent. In addition, the contribution in Eq.~\eqref{eq:sghyv} originates solely from $K_0$ in Eq.~\eqref{eq:sghy} since $K = 3 a^{\prime}/a$ vanishes at $r = 0$; however, there is no need to subtract $K_0$ from $K$ on this boundary as there is no divergence of the action at $r = 0$. If we just use $K$ here instead of $K - K_0$ in Eq.~\eqref{eq:sghy}, we obtain $S_{\mathrm{tot}} = S$. Considering this ambiguity, in the following analysis, we show both the cases with and without the GHY term and regard the difference between them as theoretical uncertainty. 

\begin{figure}
    {\includegraphics[clip, width = 0.45 \textwidth]{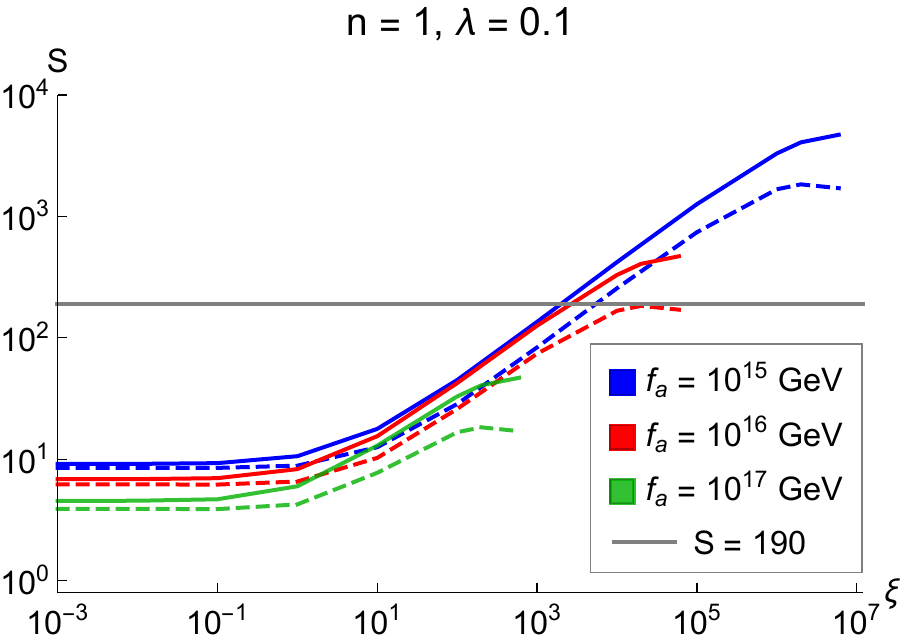}} 
    \caption{
      The values of the action without and with the GHY term as functions of $\xi$ in the solid and dashed lines, respectively. The horizontal gray line indicates the border to solve the quality problem, $S \gtrsim 190$~\cite{Kallosh:1995hi}. 
    }
    \label{fig:action-without-GHY}
\end{figure}

\begin{figure}
  {\includegraphics[clip, width = 0.45 \textwidth]{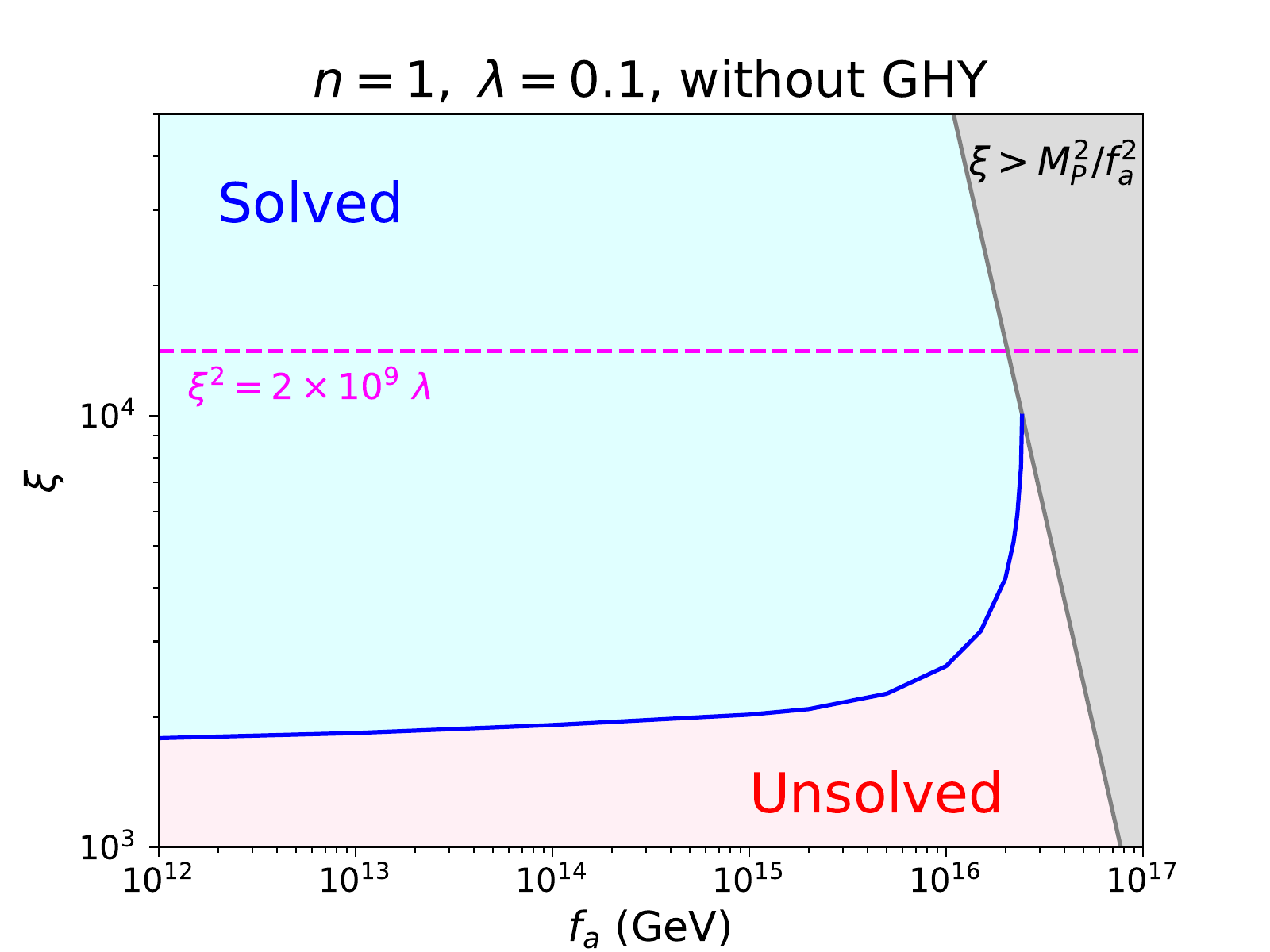}} 
  \caption{
    The region in the $f_a$--$\xi$ plane where the quality problem is avoided is shown in the blue area for the wormhole action without the GHY term. The gray region is excluded as $\xi > M_P^2/f_a^2$. 
    }  
  \label{fig:xi-fa}
\end{figure}

In Fig.~\ref{fig:action-without-GHY}, we plot the values of the action without and with the GHY term as functions of $\xi$ in the solid and dashed lines, respectively, for several choices of $f_a$ with $n = 1$ and $\lambda = 0.1$. The range of $\xi$ is limited to $\xi \leq M_P^2/f_a^2$ as we mentioned above. This figure shows that the value of the action significantly increases as $\xi$ increases. We also show in the horizontal gray line the border to solve the quality problem, $S \gtrsim 190$~\cite{Kallosh:1995hi}. As we see, the quality problem can be avoided for $\xi \gtrsim 2 \times 10^3$. In Fig.~\ref{fig:xi-fa}, we show the region where the quality problem is evaded in the blue area in the $f_a$--$\xi$ plane; we again set $n = 1$ and $\lambda = 0.1$. The gray region is excluded as $\xi > M_P^2/f_a^2$. It is found that there exists a value of $\xi$ that can solve the quality problem for $f_a \lesssim 2.5 \times 10^{16}$~GeV.
The boundary (the blue curve) remains almost constant, $\xi\sim 2\times 10^3$, for lower values of $f_a$, $f_a\lesssim 10^{12}$~GeV, which is favored in order to avoid an axion over-abundance (see, e.g., a recent work~\cite{Ballesteros:2021bee}).

It is worth noting that for $\xi = M_P^2/f_a^2$, which corresponds to the induced-gravity model, $f (r) = f_a$ is found to be the solution of Eqs.~\eqref{eq:eomf}\eqref{eq:eommett}. With this, Eq.~\eqref{eq:eommett} leads to
\begin{equation}
  a^{\prime 2} = 1 - \frac{a^4(0)}{a^4} \quad \text{with} 
  \quad a(0) = \biggl(\frac{n^2}{24\pi^4 f_a^2 M_P^2}\biggr)^{\frac{1}{4}} ~,
  \label{eq:GS_sol}
\end{equation}
which is the same as the equation that determines the Giddings-Strominger wormhole configuration~\cite{Giddings:1987cg}. In this case, we have 
\begin{equation}
  S = \sqrt{\frac{3 \pi^2}{8}} \frac{n M_P}{f_a} ~,
\end{equation}
where $S \gtrsim 190$ for $n = 1$ and $f_a \lesssim 2.5 \times 10^{16}$~GeV. This corresponds to the maximal value of $f_a$ in the blue region in Fig.~\ref{fig:xi-fa}. The parameter $\xi$, therefore, smoothly connects the minimal coupling model ($\xi = 0$), in which axion suffers from the quality problem, to the Giddings-Strominger solution ($\xi = M_P^2/f_a^2$), where the quality problem is avoided for $f_a \lesssim 2.5 \times 10^{16}$~GeV. 

Intriguingly, the size of $\xi$ in the blue region in Fig.~\ref{fig:xi-fa} is also motivated by inflation.\footnote{We, however, note that a large value of $\xi$ may indicate that the cut-off scale of the theory is as low as $\mathcal{O}(M_P/\xi)$~\cite{Burgess:2009ea, Barbon:2009ya, Burgess:2010zq,Hertzberg:2010dc}; for example, the perturbative unitarity condition for the tree-level scattering processes of $\Phi$ is violated when the center-of-mass energy is $\gtrsim M_P/\xi$. 
Such a limit imposed by the unitarity condition may be significantly relaxed if the scalar field has a large field value~\cite{Bezrukov:2010jz, Antoniadis:2021axu} (see also Refs.~\cite{Aydemir:2012nz, Calmet:2013hia}), though it is controversial~\cite{Barbon:2015fla}. 
A large $\xi$ can also cause a unitarity violation at the preheating stage~\cite{Ema:2016dny,Sfakianakis:2018lzf,Ema:2021xhq}. Other considerations based on the renormalization group analysis or the investigation of the scattering amplitudes of $\Phi$ suggest that a sizable $R^2$ term should be induced for a large $\xi$~\cite{Salvio:2015kka, Calmet:2016fsr, Ghilencea:2018rqg, Ema:2019fdd}. This term does modify the wormhole configuration and thus changes the value of the wormhole action. We will explore models including the $R^2$ term in a forthcoming paper~\cite{HKN}. }
Inflation models with an inflaton field having a non-minimal gravitational coupling have widely been suggested in the literature~\cite{Futamase:1987ua, Salopek:1988qh, Fakir:1990eg, Makino:1991sg, Fakir:1992cg, Barvinsky:1994hx, Kaiser:1994vs, Kamenshchik:1995ib, Mukaigawa:1997nh, Libanov:1998wg, Komatsu:1999mt,
Linde:2011nh, Kaiser:2013sna, Fairbairn:2014zta, Ballesteros:2016euj, Ballesteros:2016xej, Boucenna:2017fna,  McDonough:2020gmn,Spokoiny:1984bd, Accetta:1985du, Lucchin:1985ip, Fakir:1990iu, Kaiser:1994wj, Cervantes-Cota:1995ehs, Bezrukov:2007ep},
and it is known that a successful inflation can be realized in this setup. In particular, the observed CMB data~\cite{Planck:2018jri} can be explained for $\xi^2 \simeq 2 \times 10^9 \lambda$, which is shown in the pink dashed line in Fig.~\ref{fig:xi-fa} as an eye guide. The result in our work may thus motivate the scenarios in which the Peccei-Quinn field also plays the role of inflaton, as discussed in Refs.~\cite{Fairbairn:2014zta, Ballesteros:2016euj, Ballesteros:2016xej, Boucenna:2017fna}.  

All in all, the non-minimal gravitational coupling may offer a new, minimal way for axion models to evade the quality problem.

\section{Conclusion and discussions}

We have studied the effect of the non-minimal gravitational coupling of the Peccei-Quinn field on the axionic wormhole solutions. It is found that the non-minimal coupling significantly affects the value of the wormhole action and, in particular, the axion quality problem can be evaded for $\xi \gtrsim 2 \times 10^3$\footnote{The choice of $\xi$ without any reason may appear to be unnatural. The fact that the quality problem and the inflation can be explained simultaneously may justify such a choice of $\xi$.} and $f_a \lesssim 2.5 \times 10^{16}$~GeV. Interestingly, this size of the non-minimal coupling allows the field $\Phi$ to play a role of inflaton as well and to explain the observed CMB data~\cite{Planck:2018jri}. 

In the present study, we have focused on obtaining semiclassical wormhole solutions of the Euclidean functional integral. In this context, it is worth noting that several previous studies show that the Giddings-Strominger wormhole solution is in fact unstable against small fluctuations~\cite{Rubakov:1996cn, Rubakov:1996br, Kim:1997dm, Hertog:2018kbz} (see, however, Ref.~\cite{Alonso:2017avz}), indicating that this wormhole solution does not contribute to the Euclidean path integral in a conventional manner. 
It is certainly worthwhile to investigate the stability of the wormholes in our setup. If they are stable, our conclusions do not change. If unstable, it is not clear whether the quality problem exists (cf.~\cite{Hertog:2018kbz}), and whether the non-minimal coupling may relax it. We defer these issues to a future study~\cite{HKN}.

We finally note that, although we have focused on the (QCD) axion in this work, we expect similar consequences for other models based on a global U(1) symmetry, such as axion-like particles, majorons, quintessence models, relaxion models, and so on. For instance, we may consider an axion-like dark matter model in which the dark matter acquires a mass only via the wormhole effect; in this case, due to the large value of the wormhole action for a sizable $\xi$, we expect an extremely small dark-matter mass---this scenario may thus provide a natural candidate for ultralight dark matter, such as fuzzy dark matter~\cite{Hu:2000ke}. We will discuss the implications of the wormhole effect for such models on another occasion~\cite{HKN}.

\section*{Acknowledgments}
We thank Kazunori Nakayama, Yasunori Nomura, Kantaro Ohmori, and Pablo Soler for valuable discussions. This work is supported in part by the Grant-in-Aid for Innovative Areas (No.19H05810 [KH], No.19H05802 [KH], No.18H05542 [NN]), Scientific Research B (No.20H01897 [KH and NN]), Young Scientists (No.21K13916 [NN]), and JSPS Fellows (No.21J20445 [YK]).


\bibliography{ref}

\begin{thebibliography}{134}%
\makeatletter
\providecommand \@ifxundefined [1]{%
 \@ifx{#1\undefined}
}%
\providecommand \@ifnum [1]{%
 \ifnum #1\expandafter \@firstoftwo
 \else \expandafter \@secondoftwo
 \fi
}%
\providecommand \@ifx [1]{%
 \ifx #1\expandafter \@firstoftwo
 \else \expandafter \@secondoftwo
 \fi
}%
\providecommand \natexlab [1]{#1}%
\providecommand \enquote  [1]{``#1''}%
\providecommand \bibnamefont  [1]{#1}%
\providecommand \bibfnamefont [1]{#1}%
\providecommand \citenamefont [1]{#1}%
\providecommand \href@noop [0]{\@secondoftwo}%
\providecommand \href [0]{\begingroup \@sanitize@url \@href}%
\providecommand \@href[1]{\@@startlink{#1}\@@href}%
\providecommand \@@href[1]{\endgroup#1\@@endlink}%
\providecommand \@sanitize@url [0]{\catcode `\\12\catcode `\$12\catcode
  `\&12\catcode `\#12\catcode `\^12\catcode `\_12\catcode `\%12\relax}%
\providecommand \@@startlink[1]{}%
\providecommand \@@endlink[0]{}%
\providecommand \url  [0]{\begingroup\@sanitize@url \@url }%
\providecommand \@url [1]{\endgroup\@href {#1}{\urlprefix }}%
\providecommand \urlprefix  [0]{URL }%
\providecommand \Eprint [0]{\href }%
\providecommand \doibase [0]{http://dx.doi.org/}%
\providecommand \selectlanguage [0]{\@gobble}%
\providecommand \bibinfo  [0]{\@secondoftwo}%
\providecommand \bibfield  [0]{\@secondoftwo}%
\providecommand \translation [1]{[#1]}%
\providecommand \BibitemOpen [0]{}%
\providecommand \bibitemStop [0]{}%
\providecommand \bibitemNoStop [0]{.\EOS\space}%
\providecommand \EOS [0]{\spacefactor3000\relax}%
\providecommand \BibitemShut  [1]{\csname bibitem#1\endcsname}%
\let\auto@bib@innerbib\@empty
\bibitem [{\citenamefont {Peccei}\ and\ \citenamefont
  {Quinn}(1977{\natexlab{a}})}]{Peccei:1977hh}%
  \BibitemOpen
  \bibfield  {author} {\bibinfo {author} {\bibfnamefont {R.~D.}\ \bibnamefont
  {Peccei}}\ and\ \bibinfo {author} {\bibfnamefont {H.~R.}\ \bibnamefont
  {Quinn}},\ }\href {\doibase 10.1103/PhysRevLett.38.1440} {\bibfield
  {journal} {\bibinfo  {journal} {Phys. Rev. Lett.}\ }\textbf {\bibinfo
  {volume} {38}},\ \bibinfo {pages} {1440} (\bibinfo {year}
  {1977}{\natexlab{a}})}\BibitemShut {NoStop}%
\bibitem [{\citenamefont {Peccei}\ and\ \citenamefont
  {Quinn}(1977{\natexlab{b}})}]{Peccei:1977ur}%
  \BibitemOpen
  \bibfield  {author} {\bibinfo {author} {\bibfnamefont {R.~D.}\ \bibnamefont
  {Peccei}}\ and\ \bibinfo {author} {\bibfnamefont {H.~R.}\ \bibnamefont
  {Quinn}},\ }\href {\doibase 10.1103/PhysRevD.16.1791} {\bibfield  {journal}
  {\bibinfo  {journal} {Phys. Rev. D}\ }\textbf {\bibinfo {volume} {16}},\
  \bibinfo {pages} {1791} (\bibinfo {year} {1977}{\natexlab{b}})}\BibitemShut
  {NoStop}%
\bibitem [{\citenamefont {Weinberg}(1978)}]{Weinberg:1977ma}%
  \BibitemOpen
  \bibfield  {author} {\bibinfo {author} {\bibfnamefont {S.}~\bibnamefont
  {Weinberg}},\ }\href {\doibase 10.1103/PhysRevLett.40.223} {\bibfield
  {journal} {\bibinfo  {journal} {Phys. Rev. Lett.}\ }\textbf {\bibinfo
  {volume} {40}},\ \bibinfo {pages} {223} (\bibinfo {year} {1978})}\BibitemShut
  {NoStop}%
\bibitem [{\citenamefont {Wilczek}(1978)}]{Wilczek:1977pj}%
  \BibitemOpen
  \bibfield  {author} {\bibinfo {author} {\bibfnamefont {F.}~\bibnamefont
  {Wilczek}},\ }\href {\doibase 10.1103/PhysRevLett.40.279} {\bibfield
  {journal} {\bibinfo  {journal} {Phys. Rev. Lett.}\ }\textbf {\bibinfo
  {volume} {40}},\ \bibinfo {pages} {279} (\bibinfo {year} {1978})}\BibitemShut
  {NoStop}%
\bibitem [{\citenamefont {Di~Luzio}\ \emph {et~al.}(2020)\citenamefont
  {Di~Luzio}, \citenamefont {Giannotti}, \citenamefont {Nardi},\ and\
  \citenamefont {Visinelli}}]{DiLuzio:2020wdo}%
  \BibitemOpen
  \bibfield  {author} {\bibinfo {author} {\bibfnamefont {L.}~\bibnamefont
  {Di~Luzio}}, \bibinfo {author} {\bibfnamefont {M.}~\bibnamefont {Giannotti}},
  \bibinfo {author} {\bibfnamefont {E.}~\bibnamefont {Nardi}}, \ and\ \bibinfo
  {author} {\bibfnamefont {L.}~\bibnamefont {Visinelli}},\ }\href {\doibase
  10.1016/j.physrep.2020.06.002} {\bibfield  {journal} {\bibinfo  {journal}
  {Phys. Rept.}\ }\textbf {\bibinfo {volume} {870}},\ \bibinfo {pages} {1}
  (\bibinfo {year} {2020})},\ \Eprint {http://arxiv.org/abs/2003.01100}
  {arXiv:2003.01100 [hep-ph]} \BibitemShut {NoStop}%
\bibitem [{\citenamefont {Sikivie}(2021)}]{Sikivie:2020zpn}%
  \BibitemOpen
  \bibfield  {author} {\bibinfo {author} {\bibfnamefont {P.}~\bibnamefont
  {Sikivie}},\ }\href {\doibase 10.1103/RevModPhys.93.015004} {\bibfield
  {journal} {\bibinfo  {journal} {Rev. Mod. Phys.}\ }\textbf {\bibinfo {volume}
  {93}},\ \bibinfo {pages} {015004} (\bibinfo {year} {2021})},\ \Eprint
  {http://arxiv.org/abs/2003.02206} {arXiv:2003.02206 [hep-ph]} \BibitemShut
  {NoStop}%
\bibitem [{\citenamefont {Agrawal}\ \emph {et~al.}(2021)\citenamefont {Agrawal}
  \emph {et~al.}}]{Agrawal:2021dbo}%
  \BibitemOpen
  \bibfield  {author} {\bibinfo {author} {\bibfnamefont {P.}~\bibnamefont
  {Agrawal}} \emph {et~al.},\ }\href@noop {} {\  (\bibinfo {year} {2021})},\
  \Eprint {http://arxiv.org/abs/2102.12143} {arXiv:2102.12143 [hep-ph]}
  \BibitemShut {NoStop}%
\bibitem [{\citenamefont {Banks}\ and\ \citenamefont
  {Seiberg}(2011)}]{Banks:2010zn}%
  \BibitemOpen
  \bibfield  {author} {\bibinfo {author} {\bibfnamefont {T.}~\bibnamefont
  {Banks}}\ and\ \bibinfo {author} {\bibfnamefont {N.}~\bibnamefont
  {Seiberg}},\ }\href {\doibase 10.1103/PhysRevD.83.084019} {\bibfield
  {journal} {\bibinfo  {journal} {Phys. Rev. D}\ }\textbf {\bibinfo {volume}
  {83}},\ \bibinfo {pages} {084019} (\bibinfo {year} {2011})},\ \Eprint
  {http://arxiv.org/abs/1011.5120} {arXiv:1011.5120 [hep-th]} \BibitemShut
  {NoStop}%
\bibitem [{\citenamefont {Witten}(2018)}]{Witten:2017hdv}%
  \BibitemOpen
  \bibfield  {author} {\bibinfo {author} {\bibfnamefont {E.}~\bibnamefont
  {Witten}},\ }\href {\doibase 10.1038/nphys4348} {\bibfield  {journal}
  {\bibinfo  {journal} {Nature Phys.}\ }\textbf {\bibinfo {volume} {14}},\
  \bibinfo {pages} {116} (\bibinfo {year} {2018})},\ \Eprint
  {http://arxiv.org/abs/1710.01791} {arXiv:1710.01791 [hep-th]} \BibitemShut
  {NoStop}%
\bibitem [{\citenamefont {Harlow}\ and\ \citenamefont
  {Ooguri}(2019)}]{Harlow:2018jwu}%
  \BibitemOpen
  \bibfield  {author} {\bibinfo {author} {\bibfnamefont {D.}~\bibnamefont
  {Harlow}}\ and\ \bibinfo {author} {\bibfnamefont {H.}~\bibnamefont
  {Ooguri}},\ }\href {\doibase 10.1103/PhysRevLett.122.191601} {\bibfield
  {journal} {\bibinfo  {journal} {Phys. Rev. Lett.}\ }\textbf {\bibinfo
  {volume} {122}},\ \bibinfo {pages} {191601} (\bibinfo {year} {2019})},\
  \Eprint {http://arxiv.org/abs/1810.05337} {arXiv:1810.05337 [hep-th]}
  \BibitemShut {NoStop}%
\bibitem [{\citenamefont {Harlow}\ and\ \citenamefont
  {Ooguri}(2021)}]{Harlow:2018tng}%
  \BibitemOpen
  \bibfield  {author} {\bibinfo {author} {\bibfnamefont {D.}~\bibnamefont
  {Harlow}}\ and\ \bibinfo {author} {\bibfnamefont {H.}~\bibnamefont
  {Ooguri}},\ }\href {\doibase 10.1007/s00220-021-04040-y} {\bibfield
  {journal} {\bibinfo  {journal} {Commun. Math. Phys.}\ }\textbf {\bibinfo
  {volume} {383}},\ \bibinfo {pages} {1669} (\bibinfo {year} {2021})},\ \Eprint
  {http://arxiv.org/abs/1810.05338} {arXiv:1810.05338 [hep-th]} \BibitemShut
  {NoStop}%
\bibitem [{\citenamefont {Georgi}\ \emph {et~al.}(1981)\citenamefont {Georgi},
  \citenamefont {Hall},\ and\ \citenamefont {Wise}}]{Georgi:1981pu}%
  \BibitemOpen
  \bibfield  {author} {\bibinfo {author} {\bibfnamefont {H.~M.}\ \bibnamefont
  {Georgi}}, \bibinfo {author} {\bibfnamefont {L.~J.}\ \bibnamefont {Hall}}, \
  and\ \bibinfo {author} {\bibfnamefont {M.~B.}\ \bibnamefont {Wise}},\ }\href
  {\doibase 10.1016/0550-3213(81)90433-8} {\bibfield  {journal} {\bibinfo
  {journal} {Nucl. Phys. B}\ }\textbf {\bibinfo {volume} {192}},\ \bibinfo
  {pages} {409} (\bibinfo {year} {1981})}\BibitemShut {NoStop}%
\bibitem [{\citenamefont {Dine}\ and\ \citenamefont
  {Seiberg}(1986)}]{Dine:1986bg}%
  \BibitemOpen
  \bibfield  {author} {\bibinfo {author} {\bibfnamefont {M.}~\bibnamefont
  {Dine}}\ and\ \bibinfo {author} {\bibfnamefont {N.}~\bibnamefont {Seiberg}},\
  }\href {\doibase 10.1016/0550-3213(86)90043-X} {\bibfield  {journal}
  {\bibinfo  {journal} {Nucl. Phys. B}\ }\textbf {\bibinfo {volume} {273}},\
  \bibinfo {pages} {109} (\bibinfo {year} {1986})}\BibitemShut {NoStop}%
\bibitem [{\citenamefont {Kamionkowski}\ and\ \citenamefont
  {March-Russell}(1992)}]{Kamionkowski:1992mf}%
  \BibitemOpen
  \bibfield  {author} {\bibinfo {author} {\bibfnamefont {M.}~\bibnamefont
  {Kamionkowski}}\ and\ \bibinfo {author} {\bibfnamefont {J.}~\bibnamefont
  {March-Russell}},\ }\href {\doibase 10.1016/0370-2693(92)90492-M} {\bibfield
  {journal} {\bibinfo  {journal} {Phys. Lett. B}\ }\textbf {\bibinfo {volume}
  {282}},\ \bibinfo {pages} {137} (\bibinfo {year} {1992})},\ \Eprint
  {http://arxiv.org/abs/hep-th/9202003} {arXiv:hep-th/9202003} \BibitemShut
  {NoStop}%
\bibitem [{\citenamefont {Holman}\ \emph {et~al.}(1992)\citenamefont {Holman},
  \citenamefont {Hsu}, \citenamefont {Kephart}, \citenamefont {Kolb},
  \citenamefont {Watkins},\ and\ \citenamefont {Widrow}}]{Holman:1992us}%
  \BibitemOpen
  \bibfield  {author} {\bibinfo {author} {\bibfnamefont {R.}~\bibnamefont
  {Holman}}, \bibinfo {author} {\bibfnamefont {S.~D.~H.}\ \bibnamefont {Hsu}},
  \bibinfo {author} {\bibfnamefont {T.~W.}\ \bibnamefont {Kephart}}, \bibinfo
  {author} {\bibfnamefont {E.~W.}\ \bibnamefont {Kolb}}, \bibinfo {author}
  {\bibfnamefont {R.}~\bibnamefont {Watkins}}, \ and\ \bibinfo {author}
  {\bibfnamefont {L.~M.}\ \bibnamefont {Widrow}},\ }\href {\doibase
  10.1016/0370-2693(92)90491-L} {\bibfield  {journal} {\bibinfo  {journal}
  {Phys. Lett. B}\ }\textbf {\bibinfo {volume} {282}},\ \bibinfo {pages} {132}
  (\bibinfo {year} {1992})},\ \Eprint {http://arxiv.org/abs/hep-ph/9203206}
  {arXiv:hep-ph/9203206} \BibitemShut {NoStop}%
\bibitem [{\citenamefont {Barr}\ and\ \citenamefont
  {Seckel}(1992)}]{Barr:1992qq}%
  \BibitemOpen
  \bibfield  {author} {\bibinfo {author} {\bibfnamefont {S.~M.}\ \bibnamefont
  {Barr}}\ and\ \bibinfo {author} {\bibfnamefont {D.}~\bibnamefont {Seckel}},\
  }\href {\doibase 10.1103/PhysRevD.46.539} {\bibfield  {journal} {\bibinfo
  {journal} {Phys. Rev. D}\ }\textbf {\bibinfo {volume} {46}},\ \bibinfo
  {pages} {539} (\bibinfo {year} {1992})}\BibitemShut {NoStop}%
\bibitem [{\citenamefont {Ghigna}\ \emph {et~al.}(1992)\citenamefont {Ghigna},
  \citenamefont {Lusignoli},\ and\ \citenamefont {Roncadelli}}]{Ghigna:1992iv}%
  \BibitemOpen
  \bibfield  {author} {\bibinfo {author} {\bibfnamefont {S.}~\bibnamefont
  {Ghigna}}, \bibinfo {author} {\bibfnamefont {M.}~\bibnamefont {Lusignoli}}, \
  and\ \bibinfo {author} {\bibfnamefont {M.}~\bibnamefont {Roncadelli}},\
  }\href {\doibase 10.1016/0370-2693(92)90019-Z} {\bibfield  {journal}
  {\bibinfo  {journal} {Phys. Lett. B}\ }\textbf {\bibinfo {volume} {283}},\
  \bibinfo {pages} {278} (\bibinfo {year} {1992})}\BibitemShut {NoStop}%
\bibitem [{\citenamefont {Chun}\ and\ \citenamefont
  {Lukas}(1992)}]{Chun:1992bn}%
  \BibitemOpen
  \bibfield  {author} {\bibinfo {author} {\bibfnamefont {E.~J.}\ \bibnamefont
  {Chun}}\ and\ \bibinfo {author} {\bibfnamefont {A.}~\bibnamefont {Lukas}},\
  }\href {\doibase 10.1016/0370-2693(92)91266-C} {\bibfield  {journal}
  {\bibinfo  {journal} {Phys. Lett. B}\ }\textbf {\bibinfo {volume} {297}},\
  \bibinfo {pages} {298} (\bibinfo {year} {1992})},\ \Eprint
  {http://arxiv.org/abs/hep-ph/9209208} {arXiv:hep-ph/9209208} \BibitemShut
  {NoStop}%
\bibitem [{\citenamefont {Cheng}\ and\ \citenamefont
  {Kaplan}(2001)}]{Cheng:2001ys}%
  \BibitemOpen
  \bibfield  {author} {\bibinfo {author} {\bibfnamefont {H.-C.}\ \bibnamefont
  {Cheng}}\ and\ \bibinfo {author} {\bibfnamefont {D.~E.}\ \bibnamefont
  {Kaplan}},\ }\href@noop {} {\  (\bibinfo {year} {2001})},\ \Eprint
  {http://arxiv.org/abs/hep-ph/0103346} {arXiv:hep-ph/0103346} \BibitemShut
  {NoStop}%
\bibitem [{\citenamefont {Hill}\ and\ \citenamefont
  {Leibovich}(2002)}]{Hill:2002kq}%
  \BibitemOpen
  \bibfield  {author} {\bibinfo {author} {\bibfnamefont {C.~T.}\ \bibnamefont
  {Hill}}\ and\ \bibinfo {author} {\bibfnamefont {A.~K.}\ \bibnamefont
  {Leibovich}},\ }\href {\doibase 10.1103/PhysRevD.66.075010} {\bibfield
  {journal} {\bibinfo  {journal} {Phys. Rev. D}\ }\textbf {\bibinfo {volume}
  {66}},\ \bibinfo {pages} {075010} (\bibinfo {year} {2002})},\ \Eprint
  {http://arxiv.org/abs/hep-ph/0205237} {arXiv:hep-ph/0205237} \BibitemShut
  {NoStop}%
\bibitem [{\citenamefont {Babu}\ \emph {et~al.}(2003)\citenamefont {Babu},
  \citenamefont {Gogoladze},\ and\ \citenamefont {Wang}}]{Babu:2002ic}%
  \BibitemOpen
  \bibfield  {author} {\bibinfo {author} {\bibfnamefont {K.~S.}\ \bibnamefont
  {Babu}}, \bibinfo {author} {\bibfnamefont {I.}~\bibnamefont {Gogoladze}}, \
  and\ \bibinfo {author} {\bibfnamefont {K.}~\bibnamefont {Wang}},\ }\href
  {\doibase 10.1016/S0370-2693(03)00411-8} {\bibfield  {journal} {\bibinfo
  {journal} {Phys. Lett. B}\ }\textbf {\bibinfo {volume} {560}},\ \bibinfo
  {pages} {214} (\bibinfo {year} {2003})},\ \Eprint
  {http://arxiv.org/abs/hep-ph/0212339} {arXiv:hep-ph/0212339} \BibitemShut
  {NoStop}%
\bibitem [{\citenamefont {Dias}\ \emph {et~al.}(2003)\citenamefont {Dias},
  \citenamefont {Pleitez},\ and\ \citenamefont {Tonasse}}]{Dias:2002gg}%
  \BibitemOpen
  \bibfield  {author} {\bibinfo {author} {\bibfnamefont {A.~G.}\ \bibnamefont
  {Dias}}, \bibinfo {author} {\bibfnamefont {V.}~\bibnamefont {Pleitez}}, \
  and\ \bibinfo {author} {\bibfnamefont {M.~D.}\ \bibnamefont {Tonasse}},\
  }\href {\doibase 10.1103/PhysRevD.67.095008} {\bibfield  {journal} {\bibinfo
  {journal} {Phys. Rev. D}\ }\textbf {\bibinfo {volume} {67}},\ \bibinfo
  {pages} {095008} (\bibinfo {year} {2003})},\ \Eprint
  {http://arxiv.org/abs/hep-ph/0211107} {arXiv:hep-ph/0211107} \BibitemShut
  {NoStop}%
\bibitem [{\citenamefont {Harigaya}\ \emph {et~al.}(2013)\citenamefont
  {Harigaya}, \citenamefont {Ibe}, \citenamefont {Schmitz},\ and\ \citenamefont
  {Yanagida}}]{Harigaya:2013vja}%
  \BibitemOpen
  \bibfield  {author} {\bibinfo {author} {\bibfnamefont {K.}~\bibnamefont
  {Harigaya}}, \bibinfo {author} {\bibfnamefont {M.}~\bibnamefont {Ibe}},
  \bibinfo {author} {\bibfnamefont {K.}~\bibnamefont {Schmitz}}, \ and\
  \bibinfo {author} {\bibfnamefont {T.~T.}\ \bibnamefont {Yanagida}},\ }\href
  {\doibase 10.1103/PhysRevD.88.075022} {\bibfield  {journal} {\bibinfo
  {journal} {Phys. Rev. D}\ }\textbf {\bibinfo {volume} {88}},\ \bibinfo
  {pages} {075022} (\bibinfo {year} {2013})},\ \Eprint
  {http://arxiv.org/abs/1308.1227} {arXiv:1308.1227 [hep-ph]} \BibitemShut
  {NoStop}%
\bibitem [{\citenamefont {Fukuda}\ \emph {et~al.}(2017)\citenamefont {Fukuda},
  \citenamefont {Ibe}, \citenamefont {Suzuki},\ and\ \citenamefont
  {Yanagida}}]{Fukuda:2017ylt}%
  \BibitemOpen
  \bibfield  {author} {\bibinfo {author} {\bibfnamefont {H.}~\bibnamefont
  {Fukuda}}, \bibinfo {author} {\bibfnamefont {M.}~\bibnamefont {Ibe}},
  \bibinfo {author} {\bibfnamefont {M.}~\bibnamefont {Suzuki}}, \ and\ \bibinfo
  {author} {\bibfnamefont {T.~T.}\ \bibnamefont {Yanagida}},\ }\href {\doibase
  10.1016/j.physletb.2017.05.071} {\bibfield  {journal} {\bibinfo  {journal}
  {Phys. Lett. B}\ }\textbf {\bibinfo {volume} {771}},\ \bibinfo {pages} {327}
  (\bibinfo {year} {2017})},\ \Eprint {http://arxiv.org/abs/1703.01112}
  {arXiv:1703.01112 [hep-ph]} \BibitemShut {NoStop}%
\bibitem [{\citenamefont {Di~Luzio}\ \emph {et~al.}(2017)\citenamefont
  {Di~Luzio}, \citenamefont {Nardi},\ and\ \citenamefont
  {Ubaldi}}]{DiLuzio:2017tjx}%
  \BibitemOpen
  \bibfield  {author} {\bibinfo {author} {\bibfnamefont {L.}~\bibnamefont
  {Di~Luzio}}, \bibinfo {author} {\bibfnamefont {E.}~\bibnamefont {Nardi}}, \
  and\ \bibinfo {author} {\bibfnamefont {L.}~\bibnamefont {Ubaldi}},\ }\href
  {\doibase 10.1103/PhysRevLett.119.011801} {\bibfield  {journal} {\bibinfo
  {journal} {Phys. Rev. Lett.}\ }\textbf {\bibinfo {volume} {119}},\ \bibinfo
  {pages} {011801} (\bibinfo {year} {2017})},\ \Eprint
  {http://arxiv.org/abs/1704.01122} {arXiv:1704.01122 [hep-ph]} \BibitemShut
  {NoStop}%
\bibitem [{\citenamefont {Duerr}\ \emph {et~al.}(2018)\citenamefont {Duerr},
  \citenamefont {Schmidt-Hoberg},\ and\ \citenamefont {Unwin}}]{Duerr:2017amf}%
  \BibitemOpen
  \bibfield  {author} {\bibinfo {author} {\bibfnamefont {M.}~\bibnamefont
  {Duerr}}, \bibinfo {author} {\bibfnamefont {K.}~\bibnamefont
  {Schmidt-Hoberg}}, \ and\ \bibinfo {author} {\bibfnamefont {J.}~\bibnamefont
  {Unwin}},\ }\href {\doibase 10.1016/j.physletb.2018.03.054} {\bibfield
  {journal} {\bibinfo  {journal} {Phys. Lett. B}\ }\textbf {\bibinfo {volume}
  {780}},\ \bibinfo {pages} {553} (\bibinfo {year} {2018})},\ \Eprint
  {http://arxiv.org/abs/1712.01841} {arXiv:1712.01841 [hep-ph]} \BibitemShut
  {NoStop}%
\bibitem [{\citenamefont {Fukuda}\ \emph {et~al.}(2018)\citenamefont {Fukuda},
  \citenamefont {Ibe}, \citenamefont {Suzuki},\ and\ \citenamefont
  {Yanagida}}]{Fukuda:2018oco}%
  \BibitemOpen
  \bibfield  {author} {\bibinfo {author} {\bibfnamefont {H.}~\bibnamefont
  {Fukuda}}, \bibinfo {author} {\bibfnamefont {M.}~\bibnamefont {Ibe}},
  \bibinfo {author} {\bibfnamefont {M.}~\bibnamefont {Suzuki}}, \ and\ \bibinfo
  {author} {\bibfnamefont {T.~T.}\ \bibnamefont {Yanagida}},\ }\href {\doibase
  10.1007/JHEP07(2018)128} {\bibfield  {journal} {\bibinfo  {journal} {JHEP}\
  }\textbf {\bibinfo {volume} {07}},\ \bibinfo {pages} {128} (\bibinfo {year}
  {2018})},\ \Eprint {http://arxiv.org/abs/1803.00759} {arXiv:1803.00759
  [hep-ph]} \BibitemShut {NoStop}%
\bibitem [{\citenamefont {Bonnefoy}\ \emph {et~al.}(2019)\citenamefont
  {Bonnefoy}, \citenamefont {Dudas},\ and\ \citenamefont
  {Pokorski}}]{Bonnefoy:2018ibr}%
  \BibitemOpen
  \bibfield  {author} {\bibinfo {author} {\bibfnamefont {Q.}~\bibnamefont
  {Bonnefoy}}, \bibinfo {author} {\bibfnamefont {E.}~\bibnamefont {Dudas}}, \
  and\ \bibinfo {author} {\bibfnamefont {S.}~\bibnamefont {Pokorski}},\ }\href
  {\doibase 10.1140/epjc/s10052-018-6528-z} {\bibfield  {journal} {\bibinfo
  {journal} {Eur. Phys. J. C}\ }\textbf {\bibinfo {volume} {79}},\ \bibinfo
  {pages} {31} (\bibinfo {year} {2019})},\ \Eprint
  {http://arxiv.org/abs/1804.01112} {arXiv:1804.01112 [hep-ph]} \BibitemShut
  {NoStop}%
\bibitem [{\citenamefont {Ibe}\ \emph {et~al.}(2018)\citenamefont {Ibe},
  \citenamefont {Suzuki},\ and\ \citenamefont {Yanagida}}]{Ibe:2018hir}%
  \BibitemOpen
  \bibfield  {author} {\bibinfo {author} {\bibfnamefont {M.}~\bibnamefont
  {Ibe}}, \bibinfo {author} {\bibfnamefont {M.}~\bibnamefont {Suzuki}}, \ and\
  \bibinfo {author} {\bibfnamefont {T.~T.}\ \bibnamefont {Yanagida}},\ }\href
  {\doibase 10.1007/JHEP08(2018)049} {\bibfield  {journal} {\bibinfo  {journal}
  {JHEP}\ }\textbf {\bibinfo {volume} {08}},\ \bibinfo {pages} {049} (\bibinfo
  {year} {2018})},\ \Eprint {http://arxiv.org/abs/1805.10029} {arXiv:1805.10029
  [hep-ph]} \BibitemShut {NoStop}%
\bibitem [{\citenamefont {Choi}\ \emph {et~al.}(2020)\citenamefont {Choi},
  \citenamefont {Suzuki},\ and\ \citenamefont {Yanagida}}]{Choi:2020vgb}%
  \BibitemOpen
  \bibfield  {author} {\bibinfo {author} {\bibfnamefont {G.}~\bibnamefont
  {Choi}}, \bibinfo {author} {\bibfnamefont {M.}~\bibnamefont {Suzuki}}, \ and\
  \bibinfo {author} {\bibfnamefont {T.~T.}\ \bibnamefont {Yanagida}},\ }\href
  {\doibase 10.1007/JHEP07(2020)048} {\bibfield  {journal} {\bibinfo  {journal}
  {JHEP}\ }\textbf {\bibinfo {volume} {07}},\ \bibinfo {pages} {048} (\bibinfo
  {year} {2020})},\ \Eprint {http://arxiv.org/abs/2005.10415} {arXiv:2005.10415
  [hep-ph]} \BibitemShut {NoStop}%
\bibitem [{\citenamefont {Ardu}\ \emph {et~al.}(2020)\citenamefont {Ardu},
  \citenamefont {Di~Luzio}, \citenamefont {Landini}, \citenamefont {Strumia},
  \citenamefont {Teresi},\ and\ \citenamefont {Wang}}]{Ardu:2020qmo}%
  \BibitemOpen
  \bibfield  {author} {\bibinfo {author} {\bibfnamefont {M.}~\bibnamefont
  {Ardu}}, \bibinfo {author} {\bibfnamefont {L.}~\bibnamefont {Di~Luzio}},
  \bibinfo {author} {\bibfnamefont {G.}~\bibnamefont {Landini}}, \bibinfo
  {author} {\bibfnamefont {A.}~\bibnamefont {Strumia}}, \bibinfo {author}
  {\bibfnamefont {D.}~\bibnamefont {Teresi}}, \ and\ \bibinfo {author}
  {\bibfnamefont {J.-W.}\ \bibnamefont {Wang}},\ }\href {\doibase
  10.1007/JHEP11(2020)090} {\bibfield  {journal} {\bibinfo  {journal} {JHEP}\
  }\textbf {\bibinfo {volume} {11}},\ \bibinfo {pages} {090} (\bibinfo {year}
  {2020})},\ \Eprint {http://arxiv.org/abs/2007.12663} {arXiv:2007.12663
  [hep-ph]} \BibitemShut {NoStop}%
\bibitem [{\citenamefont {Yin}(2020)}]{Yin:2020dfn}%
  \BibitemOpen
  \bibfield  {author} {\bibinfo {author} {\bibfnamefont {W.}~\bibnamefont
  {Yin}},\ }\href {\doibase 10.1007/JHEP10(2020)032} {\bibfield  {journal}
  {\bibinfo  {journal} {JHEP}\ }\textbf {\bibinfo {volume} {10}},\ \bibinfo
  {pages} {032} (\bibinfo {year} {2020})},\ \Eprint
  {http://arxiv.org/abs/2007.13320} {arXiv:2007.13320 [hep-ph]} \BibitemShut
  {NoStop}%
\bibitem [{\citenamefont {Di~Luzio}(2020)}]{DiLuzio:2020qio}%
  \BibitemOpen
  \bibfield  {author} {\bibinfo {author} {\bibfnamefont {L.}~\bibnamefont
  {Di~Luzio}},\ }\href {\doibase 10.1007/JHEP11(2020)074} {\bibfield  {journal}
  {\bibinfo  {journal} {JHEP}\ }\textbf {\bibinfo {volume} {11}},\ \bibinfo
  {pages} {074} (\bibinfo {year} {2020})},\ \Eprint
  {http://arxiv.org/abs/2008.09119} {arXiv:2008.09119 [hep-ph]} \BibitemShut
  {NoStop}%
\bibitem [{\citenamefont {Kawamura}\ and\ \citenamefont
  {Raby}(2021)}]{Kawamura:2020jzb}%
  \BibitemOpen
  \bibfield  {author} {\bibinfo {author} {\bibfnamefont {J.}~\bibnamefont
  {Kawamura}}\ and\ \bibinfo {author} {\bibfnamefont {S.}~\bibnamefont
  {Raby}},\ }\href {\doibase 10.1103/PhysRevD.103.015002} {\bibfield  {journal}
  {\bibinfo  {journal} {Phys. Rev. D}\ }\textbf {\bibinfo {volume} {103}},\
  \bibinfo {pages} {015002} (\bibinfo {year} {2021})},\ \Eprint
  {http://arxiv.org/abs/2009.04582} {arXiv:2009.04582 [hep-ph]} \BibitemShut
  {NoStop}%
\bibitem [{\citenamefont {Darm\'e}\ and\ \citenamefont
  {Nardi}(2021)}]{Darme:2021cxx}%
  \BibitemOpen
  \bibfield  {author} {\bibinfo {author} {\bibfnamefont {L.}~\bibnamefont
  {Darm\'e}}\ and\ \bibinfo {author} {\bibfnamefont {E.}~\bibnamefont
  {Nardi}},\ }\href@noop {} {\  (\bibinfo {year} {2021})},\ \Eprint
  {http://arxiv.org/abs/2102.05055} {arXiv:2102.05055 [hep-ph]} \BibitemShut
  {NoStop}%
\bibitem [{\citenamefont {Chen}\ \emph {et~al.}(2021)\citenamefont {Chen},
  \citenamefont {Liu},\ and\ \citenamefont {Teng}}]{Chen:2021haa}%
  \BibitemOpen
  \bibfield  {author} {\bibinfo {author} {\bibfnamefont {N.}~\bibnamefont
  {Chen}}, \bibinfo {author} {\bibfnamefont {Y.}~\bibnamefont {Liu}}, \ and\
  \bibinfo {author} {\bibfnamefont {Z.}~\bibnamefont {Teng}},\ }\href@noop {}
  {\  (\bibinfo {year} {2021})},\ \Eprint {http://arxiv.org/abs/2106.00223}
  {arXiv:2106.00223 [hep-ph]} \BibitemShut {NoStop}%
\bibitem [{\citenamefont {Dimopoulos}(1979)}]{Dimopoulos:1979pp}%
  \BibitemOpen
  \bibfield  {author} {\bibinfo {author} {\bibfnamefont {S.}~\bibnamefont
  {Dimopoulos}},\ }\href {\doibase 10.1016/0370-2693(79)91233-4} {\bibfield
  {journal} {\bibinfo  {journal} {Phys. Lett. B}\ }\textbf {\bibinfo {volume}
  {84}},\ \bibinfo {pages} {435} (\bibinfo {year} {1979})}\BibitemShut
  {NoStop}%
\bibitem [{\citenamefont {Tye}(1981)}]{Tye:1981zy}%
  \BibitemOpen
  \bibfield  {author} {\bibinfo {author} {\bibfnamefont {S.~H.~H.}\
  \bibnamefont {Tye}},\ }\href {\doibase 10.1103/PhysRevLett.47.1035}
  {\bibfield  {journal} {\bibinfo  {journal} {Phys. Rev. Lett.}\ }\textbf
  {\bibinfo {volume} {47}},\ \bibinfo {pages} {1035} (\bibinfo {year}
  {1981})}\BibitemShut {NoStop}%
\bibitem [{\citenamefont {Rubakov}(1997)}]{Rubakov:1997vp}%
  \BibitemOpen
  \bibfield  {author} {\bibinfo {author} {\bibfnamefont {V.~A.}\ \bibnamefont
  {Rubakov}},\ }\href {\doibase 10.1134/1.567390} {\bibfield  {journal}
  {\bibinfo  {journal} {JETP Lett.}\ }\textbf {\bibinfo {volume} {65}},\
  \bibinfo {pages} {621} (\bibinfo {year} {1997})},\ \Eprint
  {http://arxiv.org/abs/hep-ph/9703409} {arXiv:hep-ph/9703409} \BibitemShut
  {NoStop}%
\bibitem [{\citenamefont {Berezhiani}\ \emph {et~al.}(2001)\citenamefont
  {Berezhiani}, \citenamefont {Gianfagna},\ and\ \citenamefont
  {Giannotti}}]{Berezhiani:2000gh}%
  \BibitemOpen
  \bibfield  {author} {\bibinfo {author} {\bibfnamefont {Z.}~\bibnamefont
  {Berezhiani}}, \bibinfo {author} {\bibfnamefont {L.}~\bibnamefont
  {Gianfagna}}, \ and\ \bibinfo {author} {\bibfnamefont {M.}~\bibnamefont
  {Giannotti}},\ }\href {\doibase 10.1016/S0370-2693(00)01392-7} {\bibfield
  {journal} {\bibinfo  {journal} {Phys. Lett. B}\ }\textbf {\bibinfo {volume}
  {500}},\ \bibinfo {pages} {286} (\bibinfo {year} {2001})},\ \Eprint
  {http://arxiv.org/abs/hep-ph/0009290} {arXiv:hep-ph/0009290} \BibitemShut
  {NoStop}%
\bibitem [{\citenamefont {Gianfagna}\ \emph {et~al.}(2004)\citenamefont
  {Gianfagna}, \citenamefont {Giannotti},\ and\ \citenamefont
  {Nesti}}]{Gianfagna:2004je}%
  \BibitemOpen
  \bibfield  {author} {\bibinfo {author} {\bibfnamefont {L.}~\bibnamefont
  {Gianfagna}}, \bibinfo {author} {\bibfnamefont {M.}~\bibnamefont
  {Giannotti}}, \ and\ \bibinfo {author} {\bibfnamefont {F.}~\bibnamefont
  {Nesti}},\ }\href {\doibase 10.1088/1126-6708/2004/10/044} {\bibfield
  {journal} {\bibinfo  {journal} {JHEP}\ }\textbf {\bibinfo {volume} {10}},\
  \bibinfo {pages} {044} (\bibinfo {year} {2004})},\ \Eprint
  {http://arxiv.org/abs/hep-ph/0409185} {arXiv:hep-ph/0409185} \BibitemShut
  {NoStop}%
\bibitem [{\citenamefont {Hook}(2015)}]{Hook:2014cda}%
  \BibitemOpen
  \bibfield  {author} {\bibinfo {author} {\bibfnamefont {A.}~\bibnamefont
  {Hook}},\ }\href {\doibase 10.1103/PhysRevLett.114.141801} {\bibfield
  {journal} {\bibinfo  {journal} {Phys. Rev. Lett.}\ }\textbf {\bibinfo
  {volume} {114}},\ \bibinfo {pages} {141801} (\bibinfo {year} {2015})},\
  \Eprint {http://arxiv.org/abs/1411.3325} {arXiv:1411.3325 [hep-ph]}
  \BibitemShut {NoStop}%
\bibitem [{\citenamefont {Fukuda}\ \emph {et~al.}(2015)\citenamefont {Fukuda},
  \citenamefont {Harigaya}, \citenamefont {Ibe},\ and\ \citenamefont
  {Yanagida}}]{Fukuda:2015ana}%
  \BibitemOpen
  \bibfield  {author} {\bibinfo {author} {\bibfnamefont {H.}~\bibnamefont
  {Fukuda}}, \bibinfo {author} {\bibfnamefont {K.}~\bibnamefont {Harigaya}},
  \bibinfo {author} {\bibfnamefont {M.}~\bibnamefont {Ibe}}, \ and\ \bibinfo
  {author} {\bibfnamefont {T.~T.}\ \bibnamefont {Yanagida}},\ }\href {\doibase
  10.1103/PhysRevD.92.015021} {\bibfield  {journal} {\bibinfo  {journal} {Phys.
  Rev. D}\ }\textbf {\bibinfo {volume} {92}},\ \bibinfo {pages} {015021}
  (\bibinfo {year} {2015})},\ \Eprint {http://arxiv.org/abs/1504.06084}
  {arXiv:1504.06084 [hep-ph]} \BibitemShut {NoStop}%
\bibitem [{\citenamefont {Albaid}\ \emph {et~al.}(2015)\citenamefont {Albaid},
  \citenamefont {Dine},\ and\ \citenamefont {Draper}}]{Albaid:2015axa}%
  \BibitemOpen
  \bibfield  {author} {\bibinfo {author} {\bibfnamefont {A.}~\bibnamefont
  {Albaid}}, \bibinfo {author} {\bibfnamefont {M.}~\bibnamefont {Dine}}, \ and\
  \bibinfo {author} {\bibfnamefont {P.}~\bibnamefont {Draper}},\ }\href
  {\doibase 10.1007/JHEP12(2015)046} {\bibfield  {journal} {\bibinfo  {journal}
  {JHEP}\ }\textbf {\bibinfo {volume} {12}},\ \bibinfo {pages} {046} (\bibinfo
  {year} {2015})},\ \Eprint {http://arxiv.org/abs/1510.03392} {arXiv:1510.03392
  [hep-ph]} \BibitemShut {NoStop}%
\bibitem [{\citenamefont {Chiang}\ \emph {et~al.}(2016)\citenamefont {Chiang},
  \citenamefont {Fukuda}, \citenamefont {Ibe},\ and\ \citenamefont
  {Yanagida}}]{Chiang:2016eav}%
  \BibitemOpen
  \bibfield  {author} {\bibinfo {author} {\bibfnamefont {C.-W.}\ \bibnamefont
  {Chiang}}, \bibinfo {author} {\bibfnamefont {H.}~\bibnamefont {Fukuda}},
  \bibinfo {author} {\bibfnamefont {M.}~\bibnamefont {Ibe}}, \ and\ \bibinfo
  {author} {\bibfnamefont {T.~T.}\ \bibnamefont {Yanagida}},\ }\href {\doibase
  10.1103/PhysRevD.93.095016} {\bibfield  {journal} {\bibinfo  {journal} {Phys.
  Rev. D}\ }\textbf {\bibinfo {volume} {93}},\ \bibinfo {pages} {095016}
  (\bibinfo {year} {2016})},\ \Eprint {http://arxiv.org/abs/1602.07909}
  {arXiv:1602.07909 [hep-ph]} \BibitemShut {NoStop}%
\bibitem [{\citenamefont {Gherghetta}\ \emph {et~al.}(2016)\citenamefont
  {Gherghetta}, \citenamefont {Nagata},\ and\ \citenamefont
  {Shifman}}]{Gherghetta:2016fhp}%
  \BibitemOpen
  \bibfield  {author} {\bibinfo {author} {\bibfnamefont {T.}~\bibnamefont
  {Gherghetta}}, \bibinfo {author} {\bibfnamefont {N.}~\bibnamefont {Nagata}},
  \ and\ \bibinfo {author} {\bibfnamefont {M.}~\bibnamefont {Shifman}},\ }\href
  {\doibase 10.1103/PhysRevD.93.115010} {\bibfield  {journal} {\bibinfo
  {journal} {Phys. Rev. D}\ }\textbf {\bibinfo {volume} {93}},\ \bibinfo
  {pages} {115010} (\bibinfo {year} {2016})},\ \Eprint
  {http://arxiv.org/abs/1604.01127} {arXiv:1604.01127 [hep-ph]} \BibitemShut
  {NoStop}%
\bibitem [{\citenamefont {Dimopoulos}\ \emph {et~al.}(2016)\citenamefont
  {Dimopoulos}, \citenamefont {Hook}, \citenamefont {Huang},\ and\
  \citenamefont {Marques-Tavares}}]{Dimopoulos:2016lvn}%
  \BibitemOpen
  \bibfield  {author} {\bibinfo {author} {\bibfnamefont {S.}~\bibnamefont
  {Dimopoulos}}, \bibinfo {author} {\bibfnamefont {A.}~\bibnamefont {Hook}},
  \bibinfo {author} {\bibfnamefont {J.}~\bibnamefont {Huang}}, \ and\ \bibinfo
  {author} {\bibfnamefont {G.}~\bibnamefont {Marques-Tavares}},\ }\href
  {\doibase 10.1007/JHEP11(2016)052} {\bibfield  {journal} {\bibinfo  {journal}
  {JHEP}\ }\textbf {\bibinfo {volume} {11}},\ \bibinfo {pages} {052} (\bibinfo
  {year} {2016})},\ \Eprint {http://arxiv.org/abs/1606.03097} {arXiv:1606.03097
  [hep-ph]} \BibitemShut {NoStop}%
\bibitem [{\citenamefont {Agrawal}\ and\ \citenamefont
  {Howe}(2018)}]{Agrawal:2017ksf}%
  \BibitemOpen
  \bibfield  {author} {\bibinfo {author} {\bibfnamefont {P.}~\bibnamefont
  {Agrawal}}\ and\ \bibinfo {author} {\bibfnamefont {K.}~\bibnamefont {Howe}},\
  }\href {\doibase 10.1007/JHEP12(2018)029} {\bibfield  {journal} {\bibinfo
  {journal} {JHEP}\ }\textbf {\bibinfo {volume} {12}},\ \bibinfo {pages} {029}
  (\bibinfo {year} {2018})},\ \Eprint {http://arxiv.org/abs/1710.04213}
  {arXiv:1710.04213 [hep-ph]} \BibitemShut {NoStop}%
\bibitem [{\citenamefont {Gherghetta}\ \emph {et~al.}(2020)\citenamefont
  {Gherghetta}, \citenamefont {Khoze}, \citenamefont {Pomarol},\ and\
  \citenamefont {Shirman}}]{Gherghetta:2020keg}%
  \BibitemOpen
  \bibfield  {author} {\bibinfo {author} {\bibfnamefont {T.}~\bibnamefont
  {Gherghetta}}, \bibinfo {author} {\bibfnamefont {V.~V.}\ \bibnamefont
  {Khoze}}, \bibinfo {author} {\bibfnamefont {A.}~\bibnamefont {Pomarol}}, \
  and\ \bibinfo {author} {\bibfnamefont {Y.}~\bibnamefont {Shirman}},\ }\href
  {\doibase 10.1007/JHEP03(2020)063} {\bibfield  {journal} {\bibinfo  {journal}
  {JHEP}\ }\textbf {\bibinfo {volume} {03}},\ \bibinfo {pages} {063} (\bibinfo
  {year} {2020})},\ \Eprint {http://arxiv.org/abs/2001.05610} {arXiv:2001.05610
  [hep-ph]} \BibitemShut {NoStop}%
\bibitem [{\citenamefont {Gherghetta}\ and\ \citenamefont
  {Nguyen}(2020)}]{Gherghetta:2020ofz}%
  \BibitemOpen
  \bibfield  {author} {\bibinfo {author} {\bibfnamefont {T.}~\bibnamefont
  {Gherghetta}}\ and\ \bibinfo {author} {\bibfnamefont {M.~D.}\ \bibnamefont
  {Nguyen}},\ }\href {\doibase 10.1007/JHEP12(2020)094} {\bibfield  {journal}
  {\bibinfo  {journal} {JHEP}\ }\textbf {\bibinfo {volume} {12}},\ \bibinfo
  {pages} {094} (\bibinfo {year} {2020})},\ \Eprint
  {http://arxiv.org/abs/2007.10875} {arXiv:2007.10875 [hep-ph]} \BibitemShut
  {NoStop}%
\bibitem [{\citenamefont {Kim}(1985)}]{Kim:1984pt}%
  \BibitemOpen
  \bibfield  {author} {\bibinfo {author} {\bibfnamefont {J.~E.}\ \bibnamefont
  {Kim}},\ }\href {\doibase 10.1103/PhysRevD.31.1733} {\bibfield  {journal}
  {\bibinfo  {journal} {Phys. Rev. D}\ }\textbf {\bibinfo {volume} {31}},\
  \bibinfo {pages} {1733} (\bibinfo {year} {1985})}\BibitemShut {NoStop}%
\bibitem [{\citenamefont {Choi}\ and\ \citenamefont {Kim}(1985)}]{Choi:1985cb}%
  \BibitemOpen
  \bibfield  {author} {\bibinfo {author} {\bibfnamefont {K.}~\bibnamefont
  {Choi}}\ and\ \bibinfo {author} {\bibfnamefont {J.~E.}\ \bibnamefont {Kim}},\
  }\href {\doibase 10.1103/PhysRevD.32.1828} {\bibfield  {journal} {\bibinfo
  {journal} {Phys. Rev. D}\ }\textbf {\bibinfo {volume} {32}},\ \bibinfo
  {pages} {1828} (\bibinfo {year} {1985})}\BibitemShut {NoStop}%
\bibitem [{\citenamefont {Randall}(1992)}]{Randall:1992ut}%
  \BibitemOpen
  \bibfield  {author} {\bibinfo {author} {\bibfnamefont {L.}~\bibnamefont
  {Randall}},\ }\href {\doibase 10.1016/0370-2693(92)91928-3} {\bibfield
  {journal} {\bibinfo  {journal} {Phys. Lett. B}\ }\textbf {\bibinfo {volume}
  {284}},\ \bibinfo {pages} {77} (\bibinfo {year} {1992})}\BibitemShut
  {NoStop}%
\bibitem [{\citenamefont {Redi}\ and\ \citenamefont
  {Sato}(2016)}]{Redi:2016esr}%
  \BibitemOpen
  \bibfield  {author} {\bibinfo {author} {\bibfnamefont {M.}~\bibnamefont
  {Redi}}\ and\ \bibinfo {author} {\bibfnamefont {R.}~\bibnamefont {Sato}},\
  }\href {\doibase 10.1007/JHEP05(2016)104} {\bibfield  {journal} {\bibinfo
  {journal} {JHEP}\ }\textbf {\bibinfo {volume} {05}},\ \bibinfo {pages} {104}
  (\bibinfo {year} {2016})},\ \Eprint {http://arxiv.org/abs/1602.05427}
  {arXiv:1602.05427 [hep-ph]} \BibitemShut {NoStop}%
\bibitem [{\citenamefont {Lillard}\ and\ \citenamefont
  {Tait}(2017)}]{Lillard:2017cwx}%
  \BibitemOpen
  \bibfield  {author} {\bibinfo {author} {\bibfnamefont {B.}~\bibnamefont
  {Lillard}}\ and\ \bibinfo {author} {\bibfnamefont {T.~M.~P.}\ \bibnamefont
  {Tait}},\ }\href {\doibase 10.1007/JHEP11(2017)005} {\bibfield  {journal}
  {\bibinfo  {journal} {JHEP}\ }\textbf {\bibinfo {volume} {11}},\ \bibinfo
  {pages} {005} (\bibinfo {year} {2017})},\ \Eprint
  {http://arxiv.org/abs/1707.04261} {arXiv:1707.04261 [hep-ph]} \BibitemShut
  {NoStop}%
\bibitem [{\citenamefont {Lillard}\ and\ \citenamefont
  {Tait}(2018)}]{Lillard:2018fdt}%
  \BibitemOpen
  \bibfield  {author} {\bibinfo {author} {\bibfnamefont {B.}~\bibnamefont
  {Lillard}}\ and\ \bibinfo {author} {\bibfnamefont {T.~M.~P.}\ \bibnamefont
  {Tait}},\ }\href {\doibase 10.1007/JHEP11(2018)199} {\bibfield  {journal}
  {\bibinfo  {journal} {JHEP}\ }\textbf {\bibinfo {volume} {11}},\ \bibinfo
  {pages} {199} (\bibinfo {year} {2018})},\ \Eprint
  {http://arxiv.org/abs/1811.03089} {arXiv:1811.03089 [hep-ph]} \BibitemShut
  {NoStop}%
\bibitem [{\citenamefont {Lee}\ and\ \citenamefont {Yin}(2019)}]{Lee:2018yak}%
  \BibitemOpen
  \bibfield  {author} {\bibinfo {author} {\bibfnamefont {H.-S.}\ \bibnamefont
  {Lee}}\ and\ \bibinfo {author} {\bibfnamefont {W.}~\bibnamefont {Yin}},\
  }\href {\doibase 10.1103/PhysRevD.99.015041} {\bibfield  {journal} {\bibinfo
  {journal} {Phys. Rev. D}\ }\textbf {\bibinfo {volume} {99}},\ \bibinfo
  {pages} {015041} (\bibinfo {year} {2019})},\ \Eprint
  {http://arxiv.org/abs/1811.04039} {arXiv:1811.04039 [hep-ph]} \BibitemShut
  {NoStop}%
\bibitem [{\citenamefont {Gavela}\ \emph {et~al.}(2019)\citenamefont {Gavela},
  \citenamefont {Ibe}, \citenamefont {Quilez},\ and\ \citenamefont
  {Yanagida}}]{Gavela:2018paw}%
  \BibitemOpen
  \bibfield  {author} {\bibinfo {author} {\bibfnamefont {M.~B.}\ \bibnamefont
  {Gavela}}, \bibinfo {author} {\bibfnamefont {M.}~\bibnamefont {Ibe}},
  \bibinfo {author} {\bibfnamefont {P.}~\bibnamefont {Quilez}}, \ and\ \bibinfo
  {author} {\bibfnamefont {T.~T.}\ \bibnamefont {Yanagida}},\ }\href {\doibase
  10.1140/epjc/s10052-019-7046-3} {\bibfield  {journal} {\bibinfo  {journal}
  {Eur. Phys. J. C}\ }\textbf {\bibinfo {volume} {79}},\ \bibinfo {pages} {542}
  (\bibinfo {year} {2019})},\ \Eprint {http://arxiv.org/abs/1812.08174}
  {arXiv:1812.08174 [hep-ph]} \BibitemShut {NoStop}%
\bibitem [{\citenamefont {Ishida}\ \emph {et~al.}(2021)\citenamefont {Ishida},
  \citenamefont {Matsuzaki},\ and\ \citenamefont {Peng}}]{Ishida:2021avk}%
  \BibitemOpen
  \bibfield  {author} {\bibinfo {author} {\bibfnamefont {H.}~\bibnamefont
  {Ishida}}, \bibinfo {author} {\bibfnamefont {S.}~\bibnamefont {Matsuzaki}}, \
  and\ \bibinfo {author} {\bibfnamefont {X.-C.}\ \bibnamefont {Peng}},\
  }\href@noop {} {\  (\bibinfo {year} {2021})},\ \Eprint
  {http://arxiv.org/abs/2103.13644} {arXiv:2103.13644 [hep-ph]} \BibitemShut
  {NoStop}%
\bibitem [{\citenamefont {Carpenter}\ \emph {et~al.}(2009)\citenamefont
  {Carpenter}, \citenamefont {Dine},\ and\ \citenamefont
  {Festuccia}}]{Carpenter:2009zs}%
  \BibitemOpen
  \bibfield  {author} {\bibinfo {author} {\bibfnamefont {L.~M.}\ \bibnamefont
  {Carpenter}}, \bibinfo {author} {\bibfnamefont {M.}~\bibnamefont {Dine}}, \
  and\ \bibinfo {author} {\bibfnamefont {G.}~\bibnamefont {Festuccia}},\ }\href
  {\doibase 10.1103/PhysRevD.80.125017} {\bibfield  {journal} {\bibinfo
  {journal} {Phys. Rev. D}\ }\textbf {\bibinfo {volume} {80}},\ \bibinfo
  {pages} {125017} (\bibinfo {year} {2009})},\ \Eprint
  {http://arxiv.org/abs/0906.1273} {arXiv:0906.1273 [hep-th]} \BibitemShut
  {NoStop}%
\bibitem [{\citenamefont {Yamada}\ \emph {et~al.}(2016)\citenamefont {Yamada},
  \citenamefont {Yanagida},\ and\ \citenamefont {Yonekura}}]{Yamada:2015waa}%
  \BibitemOpen
  \bibfield  {author} {\bibinfo {author} {\bibfnamefont {M.}~\bibnamefont
  {Yamada}}, \bibinfo {author} {\bibfnamefont {T.~T.}\ \bibnamefont
  {Yanagida}}, \ and\ \bibinfo {author} {\bibfnamefont {K.}~\bibnamefont
  {Yonekura}},\ }\href {\doibase 10.1103/PhysRevLett.116.051801} {\bibfield
  {journal} {\bibinfo  {journal} {Phys. Rev. Lett.}\ }\textbf {\bibinfo
  {volume} {116}},\ \bibinfo {pages} {051801} (\bibinfo {year} {2016})},\
  \Eprint {http://arxiv.org/abs/1510.06504} {arXiv:1510.06504 [hep-ph]}
  \BibitemShut {NoStop}%
\bibitem [{\citenamefont {Higaki}\ \emph {et~al.}(2016)\citenamefont {Higaki},
  \citenamefont {Jeong}, \citenamefont {Kitajima},\ and\ \citenamefont
  {Takahashi}}]{Higaki:2016yqk}%
  \BibitemOpen
  \bibfield  {author} {\bibinfo {author} {\bibfnamefont {T.}~\bibnamefont
  {Higaki}}, \bibinfo {author} {\bibfnamefont {K.~S.}\ \bibnamefont {Jeong}},
  \bibinfo {author} {\bibfnamefont {N.}~\bibnamefont {Kitajima}}, \ and\
  \bibinfo {author} {\bibfnamefont {F.}~\bibnamefont {Takahashi}},\ }\href
  {\doibase 10.1007/JHEP06(2016)150} {\bibfield  {journal} {\bibinfo  {journal}
  {JHEP}\ }\textbf {\bibinfo {volume} {06}},\ \bibinfo {pages} {150} (\bibinfo
  {year} {2016})},\ \Eprint {http://arxiv.org/abs/1603.02090} {arXiv:1603.02090
  [hep-ph]} \BibitemShut {NoStop}%
\bibitem [{\citenamefont {Cox}\ \emph {et~al.}(2020)\citenamefont {Cox},
  \citenamefont {Gherghetta},\ and\ \citenamefont {Nguyen}}]{Cox:2019rro}%
  \BibitemOpen
  \bibfield  {author} {\bibinfo {author} {\bibfnamefont {P.}~\bibnamefont
  {Cox}}, \bibinfo {author} {\bibfnamefont {T.}~\bibnamefont {Gherghetta}}, \
  and\ \bibinfo {author} {\bibfnamefont {M.~D.}\ \bibnamefont {Nguyen}},\
  }\href {\doibase 10.1007/JHEP01(2020)188} {\bibfield  {journal} {\bibinfo
  {journal} {JHEP}\ }\textbf {\bibinfo {volume} {01}},\ \bibinfo {pages} {188}
  (\bibinfo {year} {2020})},\ \Eprint {http://arxiv.org/abs/1911.09385}
  {arXiv:1911.09385 [hep-ph]} \BibitemShut {NoStop}%
\bibitem [{\citenamefont {Bonnefoy}\ \emph {et~al.}(2021)\citenamefont
  {Bonnefoy}, \citenamefont {Cox}, \citenamefont {Dudas}, \citenamefont
  {Gherghetta},\ and\ \citenamefont {Nguyen}}]{Bonnefoy:2020llz}%
  \BibitemOpen
  \bibfield  {author} {\bibinfo {author} {\bibfnamefont {Q.}~\bibnamefont
  {Bonnefoy}}, \bibinfo {author} {\bibfnamefont {P.}~\bibnamefont {Cox}},
  \bibinfo {author} {\bibfnamefont {E.}~\bibnamefont {Dudas}}, \bibinfo
  {author} {\bibfnamefont {T.}~\bibnamefont {Gherghetta}}, \ and\ \bibinfo
  {author} {\bibfnamefont {M.~D.}\ \bibnamefont {Nguyen}},\ }\href {\doibase
  10.1007/JHEP04(2021)084} {\bibfield  {journal} {\bibinfo  {journal} {JHEP}\
  }\textbf {\bibinfo {volume} {04}},\ \bibinfo {pages} {084} (\bibinfo {year}
  {2021})},\ \Eprint {http://arxiv.org/abs/2012.09728} {arXiv:2012.09728
  [hep-ph]} \BibitemShut {NoStop}%
\bibitem [{\citenamefont {Yamada}\ and\ \citenamefont
  {Yanagida}(2021)}]{Yamada:2021uze}%
  \BibitemOpen
  \bibfield  {author} {\bibinfo {author} {\bibfnamefont {M.}~\bibnamefont
  {Yamada}}\ and\ \bibinfo {author} {\bibfnamefont {T.~T.}\ \bibnamefont
  {Yanagida}},\ }\href {\doibase 10.1016/j.physletb.2021.136267} {\bibfield
  {journal} {\bibinfo  {journal} {Phys. Lett. B}\ }\textbf {\bibinfo {volume}
  {816}},\ \bibinfo {pages} {136267} (\bibinfo {year} {2021})},\ \Eprint
  {http://arxiv.org/abs/2101.10350} {arXiv:2101.10350 [hep-ph]} \BibitemShut
  {NoStop}%
\bibitem [{\citenamefont {Di~Luzio}\ \emph {et~al.}(2021)\citenamefont
  {Di~Luzio}, \citenamefont {Gavela}, \citenamefont {Quilez},\ and\
  \citenamefont {Ringwald}}]{DiLuzio:2021pxd}%
  \BibitemOpen
  \bibfield  {author} {\bibinfo {author} {\bibfnamefont {L.}~\bibnamefont
  {Di~Luzio}}, \bibinfo {author} {\bibfnamefont {B.}~\bibnamefont {Gavela}},
  \bibinfo {author} {\bibfnamefont {P.}~\bibnamefont {Quilez}}, \ and\ \bibinfo
  {author} {\bibfnamefont {A.}~\bibnamefont {Ringwald}},\ }\href {\doibase
  10.1007/JHEP05(2021)184} {\bibfield  {journal} {\bibinfo  {journal} {JHEP}\
  }\textbf {\bibinfo {volume} {05}},\ \bibinfo {pages} {184} (\bibinfo {year}
  {2021})},\ \Eprint {http://arxiv.org/abs/2102.00012} {arXiv:2102.00012
  [hep-ph]} \BibitemShut {NoStop}%
\bibitem [{\citenamefont {Nakai}\ and\ \citenamefont
  {Suzuki}(2021)}]{Nakai:2021nyf}%
  \BibitemOpen
  \bibfield  {author} {\bibinfo {author} {\bibfnamefont {Y.}~\bibnamefont
  {Nakai}}\ and\ \bibinfo {author} {\bibfnamefont {M.}~\bibnamefont {Suzuki}},\
  }\href {\doibase 10.1016/j.physletb.2021.136239} {\bibfield  {journal}
  {\bibinfo  {journal} {Phys. Lett. B}\ }\textbf {\bibinfo {volume} {816}},\
  \bibinfo {pages} {136239} (\bibinfo {year} {2021})},\ \Eprint
  {http://arxiv.org/abs/2102.01329} {arXiv:2102.01329 [hep-ph]} \BibitemShut
  {NoStop}%
\bibitem [{\citenamefont {Bhattiprolu}\ and\ \citenamefont
  {Martin}(2021)}]{Bhattiprolu:2021rrj}%
  \BibitemOpen
  \bibfield  {author} {\bibinfo {author} {\bibfnamefont {P.~N.}\ \bibnamefont
  {Bhattiprolu}}\ and\ \bibinfo {author} {\bibfnamefont {S.~P.}\ \bibnamefont
  {Martin}},\ }\href@noop {} {\  (\bibinfo {year} {2021})},\ \Eprint
  {http://arxiv.org/abs/2106.14964} {arXiv:2106.14964 [hep-ph]} \BibitemShut
  {NoStop}%
\bibitem [{\citenamefont {Giddings}\ and\ \citenamefont
  {Strominger}(1988)}]{Giddings:1987cg}%
  \BibitemOpen
  \bibfield  {author} {\bibinfo {author} {\bibfnamefont {S.~B.}\ \bibnamefont
  {Giddings}}\ and\ \bibinfo {author} {\bibfnamefont {A.}~\bibnamefont
  {Strominger}},\ }\href {\doibase 10.1016/0550-3213(88)90446-4} {\bibfield
  {journal} {\bibinfo  {journal} {Nucl. Phys. B}\ }\textbf {\bibinfo {volume}
  {306}},\ \bibinfo {pages} {890} (\bibinfo {year} {1988})}\BibitemShut
  {NoStop}%
\bibitem [{\citenamefont {Lee}(1988)}]{Lee:1988ge}%
  \BibitemOpen
  \bibfield  {author} {\bibinfo {author} {\bibfnamefont {K.-M.}\ \bibnamefont
  {Lee}},\ }\href {\doibase 10.1103/PhysRevLett.61.263} {\bibfield  {journal}
  {\bibinfo  {journal} {Phys. Rev. Lett.}\ }\textbf {\bibinfo {volume} {61}},\
  \bibinfo {pages} {263} (\bibinfo {year} {1988})}\BibitemShut {NoStop}%
\bibitem [{\citenamefont {Rey}(1989)}]{Rey:1989mg}%
  \BibitemOpen
  \bibfield  {author} {\bibinfo {author} {\bibfnamefont {S.-J.}\ \bibnamefont
  {Rey}},\ }\href {\doibase 10.1103/PhysRevD.39.3185} {\bibfield  {journal}
  {\bibinfo  {journal} {Phys. Rev. D}\ }\textbf {\bibinfo {volume} {39}},\
  \bibinfo {pages} {3185} (\bibinfo {year} {1989})}\BibitemShut {NoStop}%
\bibitem [{\citenamefont {Abbott}\ and\ \citenamefont
  {Wise}(1989)}]{Abbott:1989jw}%
  \BibitemOpen
  \bibfield  {author} {\bibinfo {author} {\bibfnamefont {L.~F.}\ \bibnamefont
  {Abbott}}\ and\ \bibinfo {author} {\bibfnamefont {M.~B.}\ \bibnamefont
  {Wise}},\ }\href {\doibase 10.1016/0550-3213(89)90503-8} {\bibfield
  {journal} {\bibinfo  {journal} {Nucl. Phys. B}\ }\textbf {\bibinfo {volume}
  {325}},\ \bibinfo {pages} {687} (\bibinfo {year} {1989})}\BibitemShut
  {NoStop}%
\bibitem [{\citenamefont {Coleman}\ and\ \citenamefont
  {Lee}(1990)}]{Coleman:1989zu}%
  \BibitemOpen
  \bibfield  {author} {\bibinfo {author} {\bibfnamefont {S.~R.}\ \bibnamefont
  {Coleman}}\ and\ \bibinfo {author} {\bibfnamefont {K.-M.}\ \bibnamefont
  {Lee}},\ }\href {\doibase 10.1016/0550-3213(90)90149-8} {\bibfield  {journal}
  {\bibinfo  {journal} {Nucl. Phys. B}\ }\textbf {\bibinfo {volume} {329}},\
  \bibinfo {pages} {387} (\bibinfo {year} {1990})}\BibitemShut {NoStop}%
\bibitem [{\citenamefont {Hebecker}\ \emph {et~al.}(2017)\citenamefont
  {Hebecker}, \citenamefont {Mangat}, \citenamefont {Theisen},\ and\
  \citenamefont {Witkowski}}]{Hebecker:2016dsw}%
  \BibitemOpen
  \bibfield  {author} {\bibinfo {author} {\bibfnamefont {A.}~\bibnamefont
  {Hebecker}}, \bibinfo {author} {\bibfnamefont {P.}~\bibnamefont {Mangat}},
  \bibinfo {author} {\bibfnamefont {S.}~\bibnamefont {Theisen}}, \ and\
  \bibinfo {author} {\bibfnamefont {L.~T.}\ \bibnamefont {Witkowski}},\ }\href
  {\doibase 10.1007/JHEP02(2017)097} {\bibfield  {journal} {\bibinfo  {journal}
  {JHEP}\ }\textbf {\bibinfo {volume} {02}},\ \bibinfo {pages} {097} (\bibinfo
  {year} {2017})},\ \Eprint {http://arxiv.org/abs/1607.06814} {arXiv:1607.06814
  [hep-th]} \BibitemShut {NoStop}%
\bibitem [{\citenamefont {Alonso}\ and\ \citenamefont
  {Urbano}(2019)}]{Alonso:2017avz}%
  \BibitemOpen
  \bibfield  {author} {\bibinfo {author} {\bibfnamefont {R.}~\bibnamefont
  {Alonso}}\ and\ \bibinfo {author} {\bibfnamefont {A.}~\bibnamefont
  {Urbano}},\ }\href {\doibase 10.1007/JHEP02(2019)136} {\bibfield  {journal}
  {\bibinfo  {journal} {JHEP}\ }\textbf {\bibinfo {volume} {02}},\ \bibinfo
  {pages} {136} (\bibinfo {year} {2019})},\ \Eprint
  {http://arxiv.org/abs/1706.07415} {arXiv:1706.07415 [hep-ph]} \BibitemShut
  {NoStop}%
\bibitem [{\citenamefont {Hebecker}\ \emph {et~al.}(2018)\citenamefont
  {Hebecker}, \citenamefont {Mikhail},\ and\ \citenamefont
  {Soler}}]{Hebecker:2018ofv}%
  \BibitemOpen
  \bibfield  {author} {\bibinfo {author} {\bibfnamefont {A.}~\bibnamefont
  {Hebecker}}, \bibinfo {author} {\bibfnamefont {T.}~\bibnamefont {Mikhail}}, \
  and\ \bibinfo {author} {\bibfnamefont {P.}~\bibnamefont {Soler}},\ }\href
  {\doibase 10.3389/fspas.2018.00035} {\bibfield  {journal} {\bibinfo
  {journal} {Front. Astron. Space Sci.}\ }\textbf {\bibinfo {volume} {5}},\
  \bibinfo {pages} {35} (\bibinfo {year} {2018})},\ \Eprint
  {http://arxiv.org/abs/1807.00824} {arXiv:1807.00824 [hep-th]} \BibitemShut
  {NoStop}%
\bibitem [{\citenamefont {Alvey}\ and\ \citenamefont
  {Escudero}(2021)}]{Alvey:2020nyh}%
  \BibitemOpen
  \bibfield  {author} {\bibinfo {author} {\bibfnamefont {J.}~\bibnamefont
  {Alvey}}\ and\ \bibinfo {author} {\bibfnamefont {M.}~\bibnamefont
  {Escudero}},\ }\href {\doibase 10.1007/JHEP01(2021)032} {\bibfield  {journal}
  {\bibinfo  {journal} {JHEP}\ }\textbf {\bibinfo {volume} {01}},\ \bibinfo
  {pages} {032} (\bibinfo {year} {2021})},\ \Eprint
  {http://arxiv.org/abs/2009.03917} {arXiv:2009.03917 [hep-ph]} \BibitemShut
  {NoStop}%
\bibitem [{\citenamefont {Kallosh}\ \emph {et~al.}(1995)\citenamefont
  {Kallosh}, \citenamefont {Linde}, \citenamefont {Linde},\ and\ \citenamefont
  {Susskind}}]{Kallosh:1995hi}%
  \BibitemOpen
  \bibfield  {author} {\bibinfo {author} {\bibfnamefont {R.}~\bibnamefont
  {Kallosh}}, \bibinfo {author} {\bibfnamefont {A.~D.}\ \bibnamefont {Linde}},
  \bibinfo {author} {\bibfnamefont {D.~A.}\ \bibnamefont {Linde}}, \ and\
  \bibinfo {author} {\bibfnamefont {L.}~\bibnamefont {Susskind}},\ }\href
  {\doibase 10.1103/PhysRevD.52.912} {\bibfield  {journal} {\bibinfo  {journal}
  {Phys. Rev. D}\ }\textbf {\bibinfo {volume} {52}},\ \bibinfo {pages} {912}
  (\bibinfo {year} {1995})},\ \Eprint {http://arxiv.org/abs/hep-th/9502069}
  {arXiv:hep-th/9502069} \BibitemShut {NoStop}%
\bibitem [{\citenamefont {Futamase}\ and\ \citenamefont
  {Maeda}(1989)}]{Futamase:1987ua}%
  \BibitemOpen
  \bibfield  {author} {\bibinfo {author} {\bibfnamefont {T.}~\bibnamefont
  {Futamase}}\ and\ \bibinfo {author} {\bibfnamefont {K.-i.}\ \bibnamefont
  {Maeda}},\ }\href {\doibase 10.1103/PhysRevD.39.399} {\bibfield  {journal}
  {\bibinfo  {journal} {Phys. Rev. D}\ }\textbf {\bibinfo {volume} {39}},\
  \bibinfo {pages} {399} (\bibinfo {year} {1989})}\BibitemShut {NoStop}%
\bibitem [{\citenamefont {Salopek}\ \emph {et~al.}(1989)\citenamefont
  {Salopek}, \citenamefont {Bond},\ and\ \citenamefont
  {Bardeen}}]{Salopek:1988qh}%
  \BibitemOpen
  \bibfield  {author} {\bibinfo {author} {\bibfnamefont {D.~S.}\ \bibnamefont
  {Salopek}}, \bibinfo {author} {\bibfnamefont {J.~R.}\ \bibnamefont {Bond}}, \
  and\ \bibinfo {author} {\bibfnamefont {J.~M.}\ \bibnamefont {Bardeen}},\
  }\href {\doibase 10.1103/PhysRevD.40.1753} {\bibfield  {journal} {\bibinfo
  {journal} {Phys. Rev. D}\ }\textbf {\bibinfo {volume} {40}},\ \bibinfo
  {pages} {1753} (\bibinfo {year} {1989})}\BibitemShut {NoStop}%
\bibitem [{\citenamefont {Fakir}\ and\ \citenamefont
  {Unruh}(1990{\natexlab{a}})}]{Fakir:1990eg}%
  \BibitemOpen
  \bibfield  {author} {\bibinfo {author} {\bibfnamefont {R.}~\bibnamefont
  {Fakir}}\ and\ \bibinfo {author} {\bibfnamefont {W.~G.}\ \bibnamefont
  {Unruh}},\ }\href {\doibase 10.1103/PhysRevD.41.1783} {\bibfield  {journal}
  {\bibinfo  {journal} {Phys. Rev. D}\ }\textbf {\bibinfo {volume} {41}},\
  \bibinfo {pages} {1783} (\bibinfo {year} {1990}{\natexlab{a}})}\BibitemShut
  {NoStop}%
\bibitem [{\citenamefont {Makino}\ and\ \citenamefont
  {Sasaki}(1991)}]{Makino:1991sg}%
  \BibitemOpen
  \bibfield  {author} {\bibinfo {author} {\bibfnamefont {N.}~\bibnamefont
  {Makino}}\ and\ \bibinfo {author} {\bibfnamefont {M.}~\bibnamefont
  {Sasaki}},\ }\href {\doibase 10.1143/PTP.86.103} {\bibfield  {journal}
  {\bibinfo  {journal} {Prog. Theor. Phys.}\ }\textbf {\bibinfo {volume}
  {86}},\ \bibinfo {pages} {103} (\bibinfo {year} {1991})}\BibitemShut
  {NoStop}%
\bibitem [{\citenamefont {Fakir}\ \emph {et~al.}(1992)\citenamefont {Fakir},
  \citenamefont {Habib},\ and\ \citenamefont {Unruh}}]{Fakir:1992cg}%
  \BibitemOpen
  \bibfield  {author} {\bibinfo {author} {\bibfnamefont {R.}~\bibnamefont
  {Fakir}}, \bibinfo {author} {\bibfnamefont {S.}~\bibnamefont {Habib}}, \ and\
  \bibinfo {author} {\bibfnamefont {W.}~\bibnamefont {Unruh}},\ }\href
  {\doibase 10.1086/171591} {\bibfield  {journal} {\bibinfo  {journal}
  {Astrophys. J.}\ }\textbf {\bibinfo {volume} {394}},\ \bibinfo {pages} {396}
  (\bibinfo {year} {1992})}\BibitemShut {NoStop}%
\bibitem [{\citenamefont {Barvinsky}\ and\ \citenamefont
  {Kamenshchik}(1994)}]{Barvinsky:1994hx}%
  \BibitemOpen
  \bibfield  {author} {\bibinfo {author} {\bibfnamefont {A.~O.}\ \bibnamefont
  {Barvinsky}}\ and\ \bibinfo {author} {\bibfnamefont {A.~Y.}\ \bibnamefont
  {Kamenshchik}},\ }\href {\doibase 10.1016/0370-2693(94)91253-X} {\bibfield
  {journal} {\bibinfo  {journal} {Phys. Lett. B}\ }\textbf {\bibinfo {volume}
  {332}},\ \bibinfo {pages} {270} (\bibinfo {year} {1994})},\ \Eprint
  {http://arxiv.org/abs/gr-qc/9404062} {arXiv:gr-qc/9404062} \BibitemShut
  {NoStop}%
\bibitem [{\citenamefont {Kaiser}(1995)}]{Kaiser:1994vs}%
  \BibitemOpen
  \bibfield  {author} {\bibinfo {author} {\bibfnamefont {D.~I.}\ \bibnamefont
  {Kaiser}},\ }\href {\doibase 10.1103/PhysRevD.52.4295} {\bibfield  {journal}
  {\bibinfo  {journal} {Phys. Rev. D}\ }\textbf {\bibinfo {volume} {52}},\
  \bibinfo {pages} {4295} (\bibinfo {year} {1995})},\ \Eprint
  {http://arxiv.org/abs/astro-ph/9408044} {arXiv:astro-ph/9408044} \BibitemShut
  {NoStop}%
\bibitem [{\citenamefont {Kamenshchik}\ \emph {et~al.}(1995)\citenamefont
  {Kamenshchik}, \citenamefont {Khalatnikov},\ and\ \citenamefont
  {Toporensky}}]{Kamenshchik:1995ib}%
  \BibitemOpen
  \bibfield  {author} {\bibinfo {author} {\bibfnamefont {A.~Y.}\ \bibnamefont
  {Kamenshchik}}, \bibinfo {author} {\bibfnamefont {I.~M.}\ \bibnamefont
  {Khalatnikov}}, \ and\ \bibinfo {author} {\bibfnamefont {A.~V.}\ \bibnamefont
  {Toporensky}},\ }\href {\doibase 10.1016/0370-2693(95)00834-8} {\bibfield
  {journal} {\bibinfo  {journal} {Phys. Lett. B}\ }\textbf {\bibinfo {volume}
  {357}},\ \bibinfo {pages} {36} (\bibinfo {year} {1995})},\ \Eprint
  {http://arxiv.org/abs/gr-qc/9508034} {arXiv:gr-qc/9508034} \BibitemShut
  {NoStop}%
\bibitem [{\citenamefont {Mukaigawa}\ \emph {et~al.}(1998)\citenamefont
  {Mukaigawa}, \citenamefont {Muta},\ and\ \citenamefont
  {Odintsov}}]{Mukaigawa:1997nh}%
  \BibitemOpen
  \bibfield  {author} {\bibinfo {author} {\bibfnamefont {S.}~\bibnamefont
  {Mukaigawa}}, \bibinfo {author} {\bibfnamefont {T.}~\bibnamefont {Muta}}, \
  and\ \bibinfo {author} {\bibfnamefont {S.~D.}\ \bibnamefont {Odintsov}},\
  }\href {\doibase 10.1142/S0217751X98001396} {\bibfield  {journal} {\bibinfo
  {journal} {Int. J. Mod. Phys. A}\ }\textbf {\bibinfo {volume} {13}},\
  \bibinfo {pages} {2739} (\bibinfo {year} {1998})},\ \Eprint
  {http://arxiv.org/abs/hep-ph/9709299} {arXiv:hep-ph/9709299} \BibitemShut
  {NoStop}%
\bibitem [{\citenamefont {Libanov}\ \emph {et~al.}(1998)\citenamefont
  {Libanov}, \citenamefont {Rubakov},\ and\ \citenamefont
  {Tinyakov}}]{Libanov:1998wg}%
  \BibitemOpen
  \bibfield  {author} {\bibinfo {author} {\bibfnamefont {M.~V.}\ \bibnamefont
  {Libanov}}, \bibinfo {author} {\bibfnamefont {V.~A.}\ \bibnamefont
  {Rubakov}}, \ and\ \bibinfo {author} {\bibfnamefont {P.~G.}\ \bibnamefont
  {Tinyakov}},\ }\href {\doibase 10.1016/S0370-2693(98)01269-6} {\bibfield
  {journal} {\bibinfo  {journal} {Phys. Lett. B}\ }\textbf {\bibinfo {volume}
  {442}},\ \bibinfo {pages} {63} (\bibinfo {year} {1998})},\ \Eprint
  {http://arxiv.org/abs/hep-ph/9807553} {arXiv:hep-ph/9807553} \BibitemShut
  {NoStop}%
\bibitem [{\citenamefont {Komatsu}\ and\ \citenamefont
  {Futamase}(1999)}]{Komatsu:1999mt}%
  \BibitemOpen
  \bibfield  {author} {\bibinfo {author} {\bibfnamefont {E.}~\bibnamefont
  {Komatsu}}\ and\ \bibinfo {author} {\bibfnamefont {T.}~\bibnamefont
  {Futamase}},\ }\href {\doibase 10.1103/PhysRevD.59.064029} {\bibfield
  {journal} {\bibinfo  {journal} {Phys. Rev. D}\ }\textbf {\bibinfo {volume}
  {59}},\ \bibinfo {pages} {064029} (\bibinfo {year} {1999})},\ \Eprint
  {http://arxiv.org/abs/astro-ph/9901127} {arXiv:astro-ph/9901127} \BibitemShut
  {NoStop}%
\bibitem [{\citenamefont {Linde}\ \emph {et~al.}(2011)\citenamefont {Linde},
  \citenamefont {Noorbala},\ and\ \citenamefont {Westphal}}]{Linde:2011nh}%
  \BibitemOpen
  \bibfield  {author} {\bibinfo {author} {\bibfnamefont {A.}~\bibnamefont
  {Linde}}, \bibinfo {author} {\bibfnamefont {M.}~\bibnamefont {Noorbala}}, \
  and\ \bibinfo {author} {\bibfnamefont {A.}~\bibnamefont {Westphal}},\ }\href
  {\doibase 10.1088/1475-7516/2011/03/013} {\bibfield  {journal} {\bibinfo
  {journal} {JCAP}\ }\textbf {\bibinfo {volume} {03}},\ \bibinfo {pages} {013}
  (\bibinfo {year} {2011})},\ \Eprint {http://arxiv.org/abs/1101.2652}
  {arXiv:1101.2652 [hep-th]} \BibitemShut {NoStop}%
\bibitem [{\citenamefont {Kaiser}\ and\ \citenamefont
  {Sfakianakis}(2014)}]{Kaiser:2013sna}%
  \BibitemOpen
  \bibfield  {author} {\bibinfo {author} {\bibfnamefont {D.~I.}\ \bibnamefont
  {Kaiser}}\ and\ \bibinfo {author} {\bibfnamefont {E.~I.}\ \bibnamefont
  {Sfakianakis}},\ }\href {\doibase 10.1103/PhysRevLett.112.011302} {\bibfield
  {journal} {\bibinfo  {journal} {Phys. Rev. Lett.}\ }\textbf {\bibinfo
  {volume} {112}},\ \bibinfo {pages} {011302} (\bibinfo {year} {2014})},\
  \Eprint {http://arxiv.org/abs/1304.0363} {arXiv:1304.0363 [astro-ph.CO]}
  \BibitemShut {NoStop}%
\bibitem [{\citenamefont {Fairbairn}\ \emph {et~al.}(2015)\citenamefont
  {Fairbairn}, \citenamefont {Hogan},\ and\ \citenamefont
  {Marsh}}]{Fairbairn:2014zta}%
  \BibitemOpen
  \bibfield  {author} {\bibinfo {author} {\bibfnamefont {M.}~\bibnamefont
  {Fairbairn}}, \bibinfo {author} {\bibfnamefont {R.}~\bibnamefont {Hogan}}, \
  and\ \bibinfo {author} {\bibfnamefont {D.~J.~E.}\ \bibnamefont {Marsh}},\
  }\href {\doibase 10.1103/PhysRevD.91.023509} {\bibfield  {journal} {\bibinfo
  {journal} {Phys. Rev. D}\ }\textbf {\bibinfo {volume} {91}},\ \bibinfo
  {pages} {023509} (\bibinfo {year} {2015})},\ \Eprint
  {http://arxiv.org/abs/1410.1752} {arXiv:1410.1752 [hep-ph]} \BibitemShut
  {NoStop}%
\bibitem [{\citenamefont {Ballesteros}\ \emph
  {et~al.}(2017{\natexlab{a}})\citenamefont {Ballesteros}, \citenamefont
  {Redondo}, \citenamefont {Ringwald},\ and\ \citenamefont
  {Tamarit}}]{Ballesteros:2016euj}%
  \BibitemOpen
  \bibfield  {author} {\bibinfo {author} {\bibfnamefont {G.}~\bibnamefont
  {Ballesteros}}, \bibinfo {author} {\bibfnamefont {J.}~\bibnamefont
  {Redondo}}, \bibinfo {author} {\bibfnamefont {A.}~\bibnamefont {Ringwald}}, \
  and\ \bibinfo {author} {\bibfnamefont {C.}~\bibnamefont {Tamarit}},\ }\href
  {\doibase 10.1103/PhysRevLett.118.071802} {\bibfield  {journal} {\bibinfo
  {journal} {Phys. Rev. Lett.}\ }\textbf {\bibinfo {volume} {118}},\ \bibinfo
  {pages} {071802} (\bibinfo {year} {2017}{\natexlab{a}})},\ \Eprint
  {http://arxiv.org/abs/1608.05414} {arXiv:1608.05414 [hep-ph]} \BibitemShut
  {NoStop}%
\bibitem [{\citenamefont {Ballesteros}\ \emph
  {et~al.}(2017{\natexlab{b}})\citenamefont {Ballesteros}, \citenamefont
  {Redondo}, \citenamefont {Ringwald},\ and\ \citenamefont
  {Tamarit}}]{Ballesteros:2016xej}%
  \BibitemOpen
  \bibfield  {author} {\bibinfo {author} {\bibfnamefont {G.}~\bibnamefont
  {Ballesteros}}, \bibinfo {author} {\bibfnamefont {J.}~\bibnamefont
  {Redondo}}, \bibinfo {author} {\bibfnamefont {A.}~\bibnamefont {Ringwald}}, \
  and\ \bibinfo {author} {\bibfnamefont {C.}~\bibnamefont {Tamarit}},\ }\href
  {\doibase 10.1088/1475-7516/2017/08/001} {\bibfield  {journal} {\bibinfo
  {journal} {JCAP}\ }\textbf {\bibinfo {volume} {08}},\ \bibinfo {pages} {001}
  (\bibinfo {year} {2017}{\natexlab{b}})},\ \Eprint
  {http://arxiv.org/abs/1610.01639} {arXiv:1610.01639 [hep-ph]} \BibitemShut
  {NoStop}%
\bibitem [{\citenamefont {Boucenna}\ and\ \citenamefont
  {Shafi}(2018)}]{Boucenna:2017fna}%
  \BibitemOpen
  \bibfield  {author} {\bibinfo {author} {\bibfnamefont {S.~M.}\ \bibnamefont
  {Boucenna}}\ and\ \bibinfo {author} {\bibfnamefont {Q.}~\bibnamefont
  {Shafi}},\ }\href {\doibase 10.1103/PhysRevD.97.075012} {\bibfield  {journal}
  {\bibinfo  {journal} {Phys. Rev. D}\ }\textbf {\bibinfo {volume} {97}},\
  \bibinfo {pages} {075012} (\bibinfo {year} {2018})},\ \Eprint
  {http://arxiv.org/abs/1712.06526} {arXiv:1712.06526 [hep-ph]} \BibitemShut
  {NoStop}%
\bibitem [{\citenamefont {McDonough}\ \emph {et~al.}(2020)\citenamefont
  {McDonough}, \citenamefont {Guth},\ and\ \citenamefont
  {Kaiser}}]{McDonough:2020gmn}%
  \BibitemOpen
  \bibfield  {author} {\bibinfo {author} {\bibfnamefont {E.}~\bibnamefont
  {McDonough}}, \bibinfo {author} {\bibfnamefont {A.~H.}\ \bibnamefont {Guth}},
  \ and\ \bibinfo {author} {\bibfnamefont {D.~I.}\ \bibnamefont {Kaiser}},\
  }\href@noop {} {\  (\bibinfo {year} {2020})},\ \Eprint
  {http://arxiv.org/abs/2010.04179} {arXiv:2010.04179 [hep-th]} \BibitemShut
  {NoStop}%
\bibitem [{\citenamefont {Coule}\ and\ \citenamefont
  {Maeda}(1990)}]{Coule:1989xu}%
  \BibitemOpen
  \bibfield  {author} {\bibinfo {author} {\bibfnamefont {D.~H.}\ \bibnamefont
  {Coule}}\ and\ \bibinfo {author} {\bibfnamefont {K.-i.}\ \bibnamefont
  {Maeda}},\ }\href {\doibase 10.1088/0264-9381/7/6/005} {\bibfield  {journal}
  {\bibinfo  {journal} {Class. Quant. Grav.}\ }\textbf {\bibinfo {volume}
  {7}},\ \bibinfo {pages} {955} (\bibinfo {year} {1990})}\BibitemShut {NoStop}%
\bibitem [{\citenamefont {Coule}(1992)}]{Coule:1992pz}%
  \BibitemOpen
  \bibfield  {author} {\bibinfo {author} {\bibfnamefont {D.~H.}\ \bibnamefont
  {Coule}},\ }\href {\doibase 10.1088/0264-9381/9/11/004} {\bibfield  {journal}
  {\bibinfo  {journal} {Class. Quant. Grav.}\ }\textbf {\bibinfo {volume}
  {9}},\ \bibinfo {pages} {2353} (\bibinfo {year} {1992})}\BibitemShut
  {NoStop}%
\bibitem [{\citenamefont {Akrami}\ \emph {et~al.}(2020)\citenamefont {Akrami}
  \emph {et~al.}}]{Planck:2018jri}%
  \BibitemOpen
  \bibfield  {author} {\bibinfo {author} {\bibfnamefont {Y.}~\bibnamefont
  {Akrami}} \emph {et~al.} (\bibinfo {collaboration} {Planck}),\ }\href
  {\doibase 10.1051/0004-6361/201833887} {\bibfield  {journal} {\bibinfo
  {journal} {Astron. Astrophys.}\ }\textbf {\bibinfo {volume} {641}},\ \bibinfo
  {pages} {A10} (\bibinfo {year} {2020})},\ \Eprint
  {http://arxiv.org/abs/1807.06211} {arXiv:1807.06211 [astro-ph.CO]}
  \BibitemShut {NoStop}%
\bibitem [{\citenamefont {Zee}(1979)}]{Zee:1978wi}%
  \BibitemOpen
  \bibfield  {author} {\bibinfo {author} {\bibfnamefont {A.}~\bibnamefont
  {Zee}},\ }\href {\doibase 10.1103/PhysRevLett.42.417} {\bibfield  {journal}
  {\bibinfo  {journal} {Phys. Rev. Lett.}\ }\textbf {\bibinfo {volume} {42}},\
  \bibinfo {pages} {417} (\bibinfo {year} {1979})}\BibitemShut {NoStop}%
\bibitem [{\citenamefont {Smolin}(1979)}]{Smolin:1979uz}%
  \BibitemOpen
  \bibfield  {author} {\bibinfo {author} {\bibfnamefont {L.}~\bibnamefont
  {Smolin}},\ }\href {\doibase 10.1016/0550-3213(79)90059-2} {\bibfield
  {journal} {\bibinfo  {journal} {Nucl. Phys. B}\ }\textbf {\bibinfo {volume}
  {160}},\ \bibinfo {pages} {253} (\bibinfo {year} {1979})}\BibitemShut
  {NoStop}%
\bibitem [{\citenamefont {Adler}(1982)}]{Adler:1982ri}%
  \BibitemOpen
  \bibfield  {author} {\bibinfo {author} {\bibfnamefont {S.~L.}\ \bibnamefont
  {Adler}},\ }\href {\doibase 10.1103/RevModPhys.54.729} {\bibfield  {journal}
  {\bibinfo  {journal} {Rev. Mod. Phys.}\ }\textbf {\bibinfo {volume} {54}},\
  \bibinfo {pages} {729} (\bibinfo {year} {1982})},\ \bibinfo {note} {[Erratum:
  Rev.Mod.Phys. 55, 837 (1983)]}\BibitemShut {NoStop}%
\bibitem [{\citenamefont {Spokoiny}(1984)}]{Spokoiny:1984bd}%
  \BibitemOpen
  \bibfield  {author} {\bibinfo {author} {\bibfnamefont {B.~L.}\ \bibnamefont
  {Spokoiny}},\ }\href {\doibase 10.1016/0370-2693(84)90587-2} {\bibfield
  {journal} {\bibinfo  {journal} {Phys. Lett. B}\ }\textbf {\bibinfo {volume}
  {147}},\ \bibinfo {pages} {39} (\bibinfo {year} {1984})}\BibitemShut
  {NoStop}%
\bibitem [{\citenamefont {Accetta}\ \emph {et~al.}(1985)\citenamefont
  {Accetta}, \citenamefont {Zoller},\ and\ \citenamefont
  {Turner}}]{Accetta:1985du}%
  \BibitemOpen
  \bibfield  {author} {\bibinfo {author} {\bibfnamefont {F.~S.}\ \bibnamefont
  {Accetta}}, \bibinfo {author} {\bibfnamefont {D.~J.}\ \bibnamefont {Zoller}},
  \ and\ \bibinfo {author} {\bibfnamefont {M.~S.}\ \bibnamefont {Turner}},\
  }\href {\doibase 10.1103/PhysRevD.31.3046} {\bibfield  {journal} {\bibinfo
  {journal} {Phys. Rev. D}\ }\textbf {\bibinfo {volume} {31}},\ \bibinfo
  {pages} {3046} (\bibinfo {year} {1985})}\BibitemShut {NoStop}%
\bibitem [{\citenamefont {Lucchin}\ \emph {et~al.}(1986)\citenamefont
  {Lucchin}, \citenamefont {Matarrese},\ and\ \citenamefont
  {Pollock}}]{Lucchin:1985ip}%
  \BibitemOpen
  \bibfield  {author} {\bibinfo {author} {\bibfnamefont {F.}~\bibnamefont
  {Lucchin}}, \bibinfo {author} {\bibfnamefont {S.}~\bibnamefont {Matarrese}},
  \ and\ \bibinfo {author} {\bibfnamefont {M.~D.}\ \bibnamefont {Pollock}},\
  }\href {\doibase 10.1016/0370-2693(86)90592-7} {\bibfield  {journal}
  {\bibinfo  {journal} {Phys. Lett. B}\ }\textbf {\bibinfo {volume} {167}},\
  \bibinfo {pages} {163} (\bibinfo {year} {1986})}\BibitemShut {NoStop}%
\bibitem [{\citenamefont {Fakir}\ and\ \citenamefont
  {Unruh}(1990{\natexlab{b}})}]{Fakir:1990iu}%
  \BibitemOpen
  \bibfield  {author} {\bibinfo {author} {\bibfnamefont {R.}~\bibnamefont
  {Fakir}}\ and\ \bibinfo {author} {\bibfnamefont {W.~G.}\ \bibnamefont
  {Unruh}},\ }\href {\doibase 10.1103/PhysRevD.41.1792} {\bibfield  {journal}
  {\bibinfo  {journal} {Phys. Rev. D}\ }\textbf {\bibinfo {volume} {41}},\
  \bibinfo {pages} {1792} (\bibinfo {year} {1990}{\natexlab{b}})}\BibitemShut
  {NoStop}%
\bibitem [{\citenamefont {Kaiser}(1994)}]{Kaiser:1994wj}%
  \BibitemOpen
  \bibfield  {author} {\bibinfo {author} {\bibfnamefont {D.~I.}\ \bibnamefont
  {Kaiser}},\ }\href {\doibase 10.1016/0370-2693(94)91292-0} {\bibfield
  {journal} {\bibinfo  {journal} {Phys. Lett. B}\ }\textbf {\bibinfo {volume}
  {340}},\ \bibinfo {pages} {23} (\bibinfo {year} {1994})},\ \Eprint
  {http://arxiv.org/abs/astro-ph/9405029} {arXiv:astro-ph/9405029} \BibitemShut
  {NoStop}%
\bibitem [{\citenamefont {Cervantes-Cota}\ and\ \citenamefont
  {Dehnen}(1995)}]{Cervantes-Cota:1995ehs}%
  \BibitemOpen
  \bibfield  {author} {\bibinfo {author} {\bibfnamefont {J.~L.}\ \bibnamefont
  {Cervantes-Cota}}\ and\ \bibinfo {author} {\bibfnamefont {H.}~\bibnamefont
  {Dehnen}},\ }\href {\doibase 10.1016/0550-3213(95)00128-X} {\bibfield
  {journal} {\bibinfo  {journal} {Nucl. Phys. B}\ }\textbf {\bibinfo {volume}
  {442}},\ \bibinfo {pages} {391} (\bibinfo {year} {1995})},\ \Eprint
  {http://arxiv.org/abs/astro-ph/9505069} {arXiv:astro-ph/9505069} \BibitemShut
  {NoStop}%
\bibitem [{\citenamefont {York}(1972)}]{York:1972sj}%
  \BibitemOpen
  \bibfield  {author} {\bibinfo {author} {\bibfnamefont {J.~W.}\ \bibnamefont
  {York}, \bibfnamefont {Jr.}},\ }\href {\doibase 10.1103/PhysRevLett.28.1082}
  {\bibfield  {journal} {\bibinfo  {journal} {Phys. Rev. Lett.}\ }\textbf
  {\bibinfo {volume} {28}},\ \bibinfo {pages} {1082} (\bibinfo {year}
  {1972})}\BibitemShut {NoStop}%
\bibitem [{\citenamefont {Gibbons}\ and\ \citenamefont
  {Hawking}(1977)}]{Gibbons:1976ue}%
  \BibitemOpen
  \bibfield  {author} {\bibinfo {author} {\bibfnamefont {G.~W.}\ \bibnamefont
  {Gibbons}}\ and\ \bibinfo {author} {\bibfnamefont {S.~W.}\ \bibnamefont
  {Hawking}},\ }\href {\doibase 10.1103/PhysRevD.15.2752} {\bibfield  {journal}
  {\bibinfo  {journal} {Phys. Rev. D}\ }\textbf {\bibinfo {volume} {15}},\
  \bibinfo {pages} {2752} (\bibinfo {year} {1977})}\BibitemShut {NoStop}%
\bibitem [{\citenamefont {Ballesteros}\ \emph {et~al.}(2021)\citenamefont
  {Ballesteros}, \citenamefont {Ringwald}, \citenamefont {Tamarit},\ and\
  \citenamefont {Welling}}]{Ballesteros:2021bee}%
  \BibitemOpen
  \bibfield  {author} {\bibinfo {author} {\bibfnamefont {G.}~\bibnamefont
  {Ballesteros}}, \bibinfo {author} {\bibfnamefont {A.}~\bibnamefont
  {Ringwald}}, \bibinfo {author} {\bibfnamefont {C.}~\bibnamefont {Tamarit}}, \
  and\ \bibinfo {author} {\bibfnamefont {Y.}~\bibnamefont {Welling}},\ }\href
  {\doibase 10.1088/1475-7516/2021/09/036} {\bibfield  {journal} {\bibinfo
  {journal} {JCAP}\ }\textbf {\bibinfo {volume} {09}},\ \bibinfo {pages} {036}
  (\bibinfo {year} {2021})},\ \Eprint {http://arxiv.org/abs/2104.13847}
  {arXiv:2104.13847 [hep-ph]} \BibitemShut {NoStop}%
\bibitem [{\citenamefont {Burgess}\ \emph {et~al.}(2009)\citenamefont
  {Burgess}, \citenamefont {Lee},\ and\ \citenamefont
  {Trott}}]{Burgess:2009ea}%
  \BibitemOpen
  \bibfield  {author} {\bibinfo {author} {\bibfnamefont {C.~P.}\ \bibnamefont
  {Burgess}}, \bibinfo {author} {\bibfnamefont {H.~M.}\ \bibnamefont {Lee}}, \
  and\ \bibinfo {author} {\bibfnamefont {M.}~\bibnamefont {Trott}},\ }\href
  {\doibase 10.1088/1126-6708/2009/09/103} {\bibfield  {journal} {\bibinfo
  {journal} {JHEP}\ }\textbf {\bibinfo {volume} {09}},\ \bibinfo {pages} {103}
  (\bibinfo {year} {2009})},\ \Eprint {http://arxiv.org/abs/0902.4465}
  {arXiv:0902.4465 [hep-ph]} \BibitemShut {NoStop}%
\bibitem [{\citenamefont {Barbon}\ and\ \citenamefont
  {Espinosa}(2009)}]{Barbon:2009ya}%
  \BibitemOpen
  \bibfield  {author} {\bibinfo {author} {\bibfnamefont {J.~L.~F.}\
  \bibnamefont {Barbon}}\ and\ \bibinfo {author} {\bibfnamefont {J.~R.}\
  \bibnamefont {Espinosa}},\ }\href {\doibase 10.1103/PhysRevD.79.081302}
  {\bibfield  {journal} {\bibinfo  {journal} {Phys. Rev. D}\ }\textbf {\bibinfo
  {volume} {79}},\ \bibinfo {pages} {081302} (\bibinfo {year} {2009})},\
  \Eprint {http://arxiv.org/abs/0903.0355} {arXiv:0903.0355 [hep-ph]}
  \BibitemShut {NoStop}%
\bibitem [{\citenamefont {Burgess}\ \emph {et~al.}(2010)\citenamefont
  {Burgess}, \citenamefont {Lee},\ and\ \citenamefont
  {Trott}}]{Burgess:2010zq}%
  \BibitemOpen
  \bibfield  {author} {\bibinfo {author} {\bibfnamefont {C.~P.}\ \bibnamefont
  {Burgess}}, \bibinfo {author} {\bibfnamefont {H.~M.}\ \bibnamefont {Lee}}, \
  and\ \bibinfo {author} {\bibfnamefont {M.}~\bibnamefont {Trott}},\ }\href
  {\doibase 10.1007/JHEP07(2010)007} {\bibfield  {journal} {\bibinfo  {journal}
  {JHEP}\ }\textbf {\bibinfo {volume} {07}},\ \bibinfo {pages} {007} (\bibinfo
  {year} {2010})},\ \Eprint {http://arxiv.org/abs/1002.2730} {arXiv:1002.2730
  [hep-ph]} \BibitemShut {NoStop}%
\bibitem [{\citenamefont {Hertzberg}(2010)}]{Hertzberg:2010dc}%
  \BibitemOpen
  \bibfield  {author} {\bibinfo {author} {\bibfnamefont {M.~P.}\ \bibnamefont
  {Hertzberg}},\ }\href {\doibase 10.1007/JHEP11(2010)023} {\bibfield
  {journal} {\bibinfo  {journal} {JHEP}\ }\textbf {\bibinfo {volume} {11}},\
  \bibinfo {pages} {023} (\bibinfo {year} {2010})},\ \Eprint
  {http://arxiv.org/abs/1002.2995} {arXiv:1002.2995 [hep-ph]} \BibitemShut
  {NoStop}%
\bibitem [{\citenamefont {Bezrukov}\ \emph {et~al.}(2011)\citenamefont
  {Bezrukov}, \citenamefont {Magnin}, \citenamefont {Shaposhnikov},\ and\
  \citenamefont {Sibiryakov}}]{Bezrukov:2010jz}%
  \BibitemOpen
  \bibfield  {author} {\bibinfo {author} {\bibfnamefont {F.}~\bibnamefont
  {Bezrukov}}, \bibinfo {author} {\bibfnamefont {A.}~\bibnamefont {Magnin}},
  \bibinfo {author} {\bibfnamefont {M.}~\bibnamefont {Shaposhnikov}}, \ and\
  \bibinfo {author} {\bibfnamefont {S.}~\bibnamefont {Sibiryakov}},\ }\href
  {\doibase 10.1007/JHEP01(2011)016} {\bibfield  {journal} {\bibinfo  {journal}
  {JHEP}\ }\textbf {\bibinfo {volume} {01}},\ \bibinfo {pages} {016} (\bibinfo
  {year} {2011})},\ \Eprint {http://arxiv.org/abs/1008.5157} {arXiv:1008.5157
  [hep-ph]} \BibitemShut {NoStop}%
\bibitem [{\citenamefont {Antoniadis}\ \emph {et~al.}(2021)\citenamefont
  {Antoniadis}, \citenamefont {Guillen},\ and\ \citenamefont
  {Tamvakis}}]{Antoniadis:2021axu}%
  \BibitemOpen
  \bibfield  {author} {\bibinfo {author} {\bibfnamefont {I.}~\bibnamefont
  {Antoniadis}}, \bibinfo {author} {\bibfnamefont {A.}~\bibnamefont {Guillen}},
  \ and\ \bibinfo {author} {\bibfnamefont {K.}~\bibnamefont {Tamvakis}},\
  }\href {\doibase 10.1007/JHEP08(2021)018} {\bibfield  {journal} {\bibinfo
  {journal} {JHEP}\ }\textbf {\bibinfo {volume} {08}},\ \bibinfo {pages} {018}
  (\bibinfo {year} {2021})},\ \Eprint {http://arxiv.org/abs/2106.09390}
  {arXiv:2106.09390 [hep-th]} \BibitemShut {NoStop}%
\bibitem [{\citenamefont {Aydemir}\ \emph {et~al.}(2012)\citenamefont
  {Aydemir}, \citenamefont {Anber},\ and\ \citenamefont
  {Donoghue}}]{Aydemir:2012nz}%
  \BibitemOpen
  \bibfield  {author} {\bibinfo {author} {\bibfnamefont {U.}~\bibnamefont
  {Aydemir}}, \bibinfo {author} {\bibfnamefont {M.~M.}\ \bibnamefont {Anber}},
  \ and\ \bibinfo {author} {\bibfnamefont {J.~F.}\ \bibnamefont {Donoghue}},\
  }\href {\doibase 10.1103/PhysRevD.86.014025} {\bibfield  {journal} {\bibinfo
  {journal} {Phys. Rev. D}\ }\textbf {\bibinfo {volume} {86}},\ \bibinfo
  {pages} {014025} (\bibinfo {year} {2012})},\ \Eprint
  {http://arxiv.org/abs/1203.5153} {arXiv:1203.5153 [hep-ph]} \BibitemShut
  {NoStop}%
\bibitem [{\citenamefont {Calmet}\ and\ \citenamefont
  {Casadio}(2014)}]{Calmet:2013hia}%
  \BibitemOpen
  \bibfield  {author} {\bibinfo {author} {\bibfnamefont {X.}~\bibnamefont
  {Calmet}}\ and\ \bibinfo {author} {\bibfnamefont {R.}~\bibnamefont
  {Casadio}},\ }\href {\doibase 10.1016/j.physletb.2014.05.008} {\bibfield
  {journal} {\bibinfo  {journal} {Phys. Lett. B}\ }\textbf {\bibinfo {volume}
  {734}},\ \bibinfo {pages} {17} (\bibinfo {year} {2014})},\ \Eprint
  {http://arxiv.org/abs/1310.7410} {arXiv:1310.7410 [hep-ph]} \BibitemShut
  {NoStop}%
\bibitem [{\citenamefont {Barbon}\ \emph {et~al.}(2015)\citenamefont {Barbon},
  \citenamefont {Casas}, \citenamefont {Elias-Miro},\ and\ \citenamefont
  {Espinosa}}]{Barbon:2015fla}%
  \BibitemOpen
  \bibfield  {author} {\bibinfo {author} {\bibfnamefont {J.~L.~F.}\
  \bibnamefont {Barbon}}, \bibinfo {author} {\bibfnamefont {J.~A.}\
  \bibnamefont {Casas}}, \bibinfo {author} {\bibfnamefont {J.}~\bibnamefont
  {Elias-Miro}}, \ and\ \bibinfo {author} {\bibfnamefont {J.~R.}\ \bibnamefont
  {Espinosa}},\ }\href {\doibase 10.1007/JHEP09(2015)027} {\bibfield  {journal}
  {\bibinfo  {journal} {JHEP}\ }\textbf {\bibinfo {volume} {09}},\ \bibinfo
  {pages} {027} (\bibinfo {year} {2015})},\ \Eprint
  {http://arxiv.org/abs/1501.02231} {arXiv:1501.02231 [hep-ph]} \BibitemShut
  {NoStop}%
\bibitem [{\citenamefont {Ema}\ \emph {et~al.}(2017)\citenamefont {Ema},
  \citenamefont {Jinno}, \citenamefont {Mukaida},\ and\ \citenamefont
  {Nakayama}}]{Ema:2016dny}%
  \BibitemOpen
  \bibfield  {author} {\bibinfo {author} {\bibfnamefont {Y.}~\bibnamefont
  {Ema}}, \bibinfo {author} {\bibfnamefont {R.}~\bibnamefont {Jinno}}, \bibinfo
  {author} {\bibfnamefont {K.}~\bibnamefont {Mukaida}}, \ and\ \bibinfo
  {author} {\bibfnamefont {K.}~\bibnamefont {Nakayama}},\ }\href {\doibase
  10.1088/1475-7516/2017/02/045} {\bibfield  {journal} {\bibinfo  {journal}
  {JCAP}\ }\textbf {\bibinfo {volume} {02}},\ \bibinfo {pages} {045} (\bibinfo
  {year} {2017})},\ \Eprint {http://arxiv.org/abs/1609.05209} {arXiv:1609.05209
  [hep-ph]} \BibitemShut {NoStop}%
\bibitem [{\citenamefont {Sfakianakis}\ and\ \citenamefont {van~de
  Vis}(2019)}]{Sfakianakis:2018lzf}%
  \BibitemOpen
  \bibfield  {author} {\bibinfo {author} {\bibfnamefont {E.~I.}\ \bibnamefont
  {Sfakianakis}}\ and\ \bibinfo {author} {\bibfnamefont {J.}~\bibnamefont
  {van~de Vis}},\ }\href {\doibase 10.1103/PhysRevD.99.083519} {\bibfield
  {journal} {\bibinfo  {journal} {Phys. Rev. D}\ }\textbf {\bibinfo {volume}
  {99}},\ \bibinfo {pages} {083519} (\bibinfo {year} {2019})},\ \Eprint
  {http://arxiv.org/abs/1810.01304} {arXiv:1810.01304 [hep-ph]} \BibitemShut
  {NoStop}%
\bibitem [{\citenamefont {Ema}\ \emph {et~al.}(2021)\citenamefont {Ema},
  \citenamefont {Jinno}, \citenamefont {Nakayama},\ and\ \citenamefont {van~de
  Vis}}]{Ema:2021xhq}%
  \BibitemOpen
  \bibfield  {author} {\bibinfo {author} {\bibfnamefont {Y.}~\bibnamefont
  {Ema}}, \bibinfo {author} {\bibfnamefont {R.}~\bibnamefont {Jinno}}, \bibinfo
  {author} {\bibfnamefont {K.}~\bibnamefont {Nakayama}}, \ and\ \bibinfo
  {author} {\bibfnamefont {J.}~\bibnamefont {van~de Vis}},\ }\href {\doibase
  10.1103/PhysRevD.103.103536} {\bibfield  {journal} {\bibinfo  {journal}
  {Phys. Rev. D}\ }\textbf {\bibinfo {volume} {103}},\ \bibinfo {pages}
  {103536} (\bibinfo {year} {2021})},\ \Eprint
  {http://arxiv.org/abs/2102.12501} {arXiv:2102.12501 [hep-ph]} \BibitemShut
  {NoStop}%
\bibitem [{\citenamefont {Salvio}\ and\ \citenamefont
  {Mazumdar}(2015)}]{Salvio:2015kka}%
  \BibitemOpen
  \bibfield  {author} {\bibinfo {author} {\bibfnamefont {A.}~\bibnamefont
  {Salvio}}\ and\ \bibinfo {author} {\bibfnamefont {A.}~\bibnamefont
  {Mazumdar}},\ }\href {\doibase 10.1016/j.physletb.2015.09.020} {\bibfield
  {journal} {\bibinfo  {journal} {Phys. Lett. B}\ }\textbf {\bibinfo {volume}
  {750}},\ \bibinfo {pages} {194} (\bibinfo {year} {2015})},\ \Eprint
  {http://arxiv.org/abs/1506.07520} {arXiv:1506.07520 [hep-ph]} \BibitemShut
  {NoStop}%
\bibitem [{\citenamefont {Calmet}\ and\ \citenamefont
  {Kuntz}(2016)}]{Calmet:2016fsr}%
  \BibitemOpen
  \bibfield  {author} {\bibinfo {author} {\bibfnamefont {X.}~\bibnamefont
  {Calmet}}\ and\ \bibinfo {author} {\bibfnamefont {I.}~\bibnamefont {Kuntz}},\
  }\href {\doibase 10.1140/epjc/s10052-016-4136-3} {\bibfield  {journal}
  {\bibinfo  {journal} {Eur. Phys. J. C}\ }\textbf {\bibinfo {volume} {76}},\
  \bibinfo {pages} {289} (\bibinfo {year} {2016})},\ \Eprint
  {http://arxiv.org/abs/1605.02236} {arXiv:1605.02236 [hep-th]} \BibitemShut
  {NoStop}%
\bibitem [{\citenamefont {Ghilencea}(2018)}]{Ghilencea:2018rqg}%
  \BibitemOpen
  \bibfield  {author} {\bibinfo {author} {\bibfnamefont {D.~M.}\ \bibnamefont
  {Ghilencea}},\ }\href {\doibase 10.1103/PhysRevD.98.103524} {\bibfield
  {journal} {\bibinfo  {journal} {Phys. Rev. D}\ }\textbf {\bibinfo {volume}
  {98}},\ \bibinfo {pages} {103524} (\bibinfo {year} {2018})},\ \Eprint
  {http://arxiv.org/abs/1807.06900} {arXiv:1807.06900 [hep-ph]} \BibitemShut
  {NoStop}%
\bibitem [{\citenamefont {Ema}(2019)}]{Ema:2019fdd}%
  \BibitemOpen
  \bibfield  {author} {\bibinfo {author} {\bibfnamefont {Y.}~\bibnamefont
  {Ema}},\ }\href {\doibase 10.1088/1475-7516/2019/09/027} {\bibfield
  {journal} {\bibinfo  {journal} {JCAP}\ }\textbf {\bibinfo {volume} {09}},\
  \bibinfo {pages} {027} (\bibinfo {year} {2019})},\ \Eprint
  {http://arxiv.org/abs/1907.00993} {arXiv:1907.00993 [hep-ph]} \BibitemShut
  {NoStop}%
\bibitem [{\citenamefont {Hamaguchi}\ \emph {et~al.}()\citenamefont
  {Hamaguchi}, \citenamefont {Kanazawa},\ and\ \citenamefont {Nagata}}]{HKN}%
  \BibitemOpen
  \bibfield  {author} {\bibinfo {author} {\bibfnamefont {K.}~\bibnamefont
  {Hamaguchi}}, \bibinfo {author} {\bibfnamefont {Y.}~\bibnamefont {Kanazawa}},
  \ and\ \bibinfo {author} {\bibfnamefont {N.}~\bibnamefont {Nagata}},\
  }\href@noop {} {}\bibinfo {note} {In preparation}\BibitemShut {NoStop}%
\bibitem [{\citenamefont {Bezrukov}\ and\ \citenamefont
  {Shaposhnikov}(2008)}]{Bezrukov:2007ep}%
  \BibitemOpen
  \bibfield  {author} {\bibinfo {author} {\bibfnamefont {F.~L.}\ \bibnamefont
  {Bezrukov}}\ and\ \bibinfo {author} {\bibfnamefont {M.}~\bibnamefont
  {Shaposhnikov}},\ }\href {\doibase 10.1016/j.physletb.2007.11.072} {\bibfield
   {journal} {\bibinfo  {journal} {Phys. Lett. B}\ }\textbf {\bibinfo {volume}
  {659}},\ \bibinfo {pages} {703} (\bibinfo {year} {2008})},\ \Eprint
  {http://arxiv.org/abs/0710.3755} {arXiv:0710.3755 [hep-th]} \BibitemShut
  {NoStop}%
\bibitem [{\citenamefont {Rubakov}\ and\ \citenamefont
  {Shvedov}(1996{\natexlab{a}})}]{Rubakov:1996cn}%
  \BibitemOpen
  \bibfield  {author} {\bibinfo {author} {\bibfnamefont {V.~A.}\ \bibnamefont
  {Rubakov}}\ and\ \bibinfo {author} {\bibfnamefont {O.~Y.}\ \bibnamefont
  {Shvedov}},\ }\href {\doibase 10.1016/0370-2693(96)00766-6} {\bibfield
  {journal} {\bibinfo  {journal} {Phys. Lett. B}\ }\textbf {\bibinfo {volume}
  {383}},\ \bibinfo {pages} {258} (\bibinfo {year} {1996}{\natexlab{a}})},\
  \Eprint {http://arxiv.org/abs/gr-qc/9604038} {arXiv:gr-qc/9604038}
  \BibitemShut {NoStop}%
\bibitem [{\citenamefont {Rubakov}\ and\ \citenamefont
  {Shvedov}(1996{\natexlab{b}})}]{Rubakov:1996br}%
  \BibitemOpen
  \bibfield  {author} {\bibinfo {author} {\bibfnamefont {V.~A.}\ \bibnamefont
  {Rubakov}}\ and\ \bibinfo {author} {\bibfnamefont {O.~Y.}\ \bibnamefont
  {Shvedov}},\ }in\ \href@noop {} {\emph {\bibinfo {booktitle} {{9th
  International Seminar on High-energy Physics}}}}\ (\bibinfo {year} {1996})\
  pp.\ \bibinfo {pages} {206--214},\ \Eprint
  {http://arxiv.org/abs/gr-qc/9608065} {arXiv:gr-qc/9608065} \BibitemShut
  {NoStop}%
\bibitem [{\citenamefont {Kim}\ \emph {et~al.}(1997)\citenamefont {Kim},
  \citenamefont {Lee},\ and\ \citenamefont {Myung}}]{Kim:1997dm}%
  \BibitemOpen
  \bibfield  {author} {\bibinfo {author} {\bibfnamefont {J.~Y.}\ \bibnamefont
  {Kim}}, \bibinfo {author} {\bibfnamefont {H.~W.}\ \bibnamefont {Lee}}, \ and\
  \bibinfo {author} {\bibfnamefont {Y.~S.}\ \bibnamefont {Myung}},\ }\href
  {\doibase 10.1103/PhysRevD.56.6684} {\bibfield  {journal} {\bibinfo
  {journal} {Phys. Rev. D}\ }\textbf {\bibinfo {volume} {56}},\ \bibinfo
  {pages} {6684} (\bibinfo {year} {1997})},\ \Eprint
  {http://arxiv.org/abs/hep-th/9701116} {arXiv:hep-th/9701116} \BibitemShut
  {NoStop}%
\bibitem [{\citenamefont {Hertog}\ \emph {et~al.}(2019)\citenamefont {Hertog},
  \citenamefont {Truijen},\ and\ \citenamefont {Van~Riet}}]{Hertog:2018kbz}%
  \BibitemOpen
  \bibfield  {author} {\bibinfo {author} {\bibfnamefont {T.}~\bibnamefont
  {Hertog}}, \bibinfo {author} {\bibfnamefont {B.}~\bibnamefont {Truijen}}, \
  and\ \bibinfo {author} {\bibfnamefont {T.}~\bibnamefont {Van~Riet}},\ }\href
  {\doibase 10.1103/PhysRevLett.123.081302} {\bibfield  {journal} {\bibinfo
  {journal} {Phys. Rev. Lett.}\ }\textbf {\bibinfo {volume} {123}},\ \bibinfo
  {pages} {081302} (\bibinfo {year} {2019})},\ \Eprint
  {http://arxiv.org/abs/1811.12690} {arXiv:1811.12690 [hep-th]} \BibitemShut
  {NoStop}%
\bibitem [{\citenamefont {Hu}\ \emph {et~al.}(2000)\citenamefont {Hu},
  \citenamefont {Barkana},\ and\ \citenamefont {Gruzinov}}]{Hu:2000ke}%
  \BibitemOpen
  \bibfield  {author} {\bibinfo {author} {\bibfnamefont {W.}~\bibnamefont
  {Hu}}, \bibinfo {author} {\bibfnamefont {R.}~\bibnamefont {Barkana}}, \ and\
  \bibinfo {author} {\bibfnamefont {A.}~\bibnamefont {Gruzinov}},\ }\href
  {\doibase 10.1103/PhysRevLett.85.1158} {\bibfield  {journal} {\bibinfo
  {journal} {Phys. Rev. Lett.}\ }\textbf {\bibinfo {volume} {85}},\ \bibinfo
  {pages} {1158} (\bibinfo {year} {2000})},\ \Eprint
  {http://arxiv.org/abs/astro-ph/0003365} {arXiv:astro-ph/0003365} \BibitemShut
  {NoStop}%
\end{thebibliography}%

\end{document}